\documentclass{emulateapj}
\usepackage{epsfig}%,natbib}

\newcommand{\kms}{\mathrm{\;km\;s^{-1}}}

\shorttitle{First results on DEEP2 galaxy groups}
\shortauthors{Brian F. Gerke et al.}

\begin{document}

\title{The DEEP2 Galaxy Redshift Survey: First results on galaxy groups}

\author{Brian F. Gerke\altaffilmark{1},
Jeffrey A. Newman\altaffilmark{2},  Marc
Davis\altaffilmark{1,3}, Christian Marinoni\altaffilmark{4},Renbin
Yan\altaffilmark{3},   
Alison L. Coil\altaffilmark{3}, Charlie Conroy\altaffilmark{3}, Michael
C. Cooper\altaffilmark{3}, 
S. M. Faber\altaffilmark{5}, Douglas 
P. Finkbeiner\altaffilmark{6}, Puragra Guhathakurta\altaffilmark{5}, Nick Kaiser\altaffilmark{7}, David C. Koo\altaffilmark{5},
Andrew C. Phillips\altaffilmark{5}, Benjamin J. Weiner\altaffilmark{8}, Christopher N. A. Willmer\altaffilmark{5}}
\altaffiltext{1}{Department of Physics, University of California,
Berkeley, CA 94720}
\altaffiltext{2}{Hubble Fellow; Institute for 
Nuclear and Particle Astrophysics, Lawrence Berkeley National Laboratory, 
Berkeley, CA 94720} 
\altaffiltext{3}{Department of Astronomy, University of California,
Berkeley, CA 94720}
\altaffiltext{4}{Brera Astronomical Observatory, via Brera 28, 20121
Milan, Italy}
\altaffiltext{5}{University of California Observatories/Lick
Observatory, Department of Astronomy and Astrophysics, University of
California, Santa Cruz, CA 95064}
\altaffiltext{6}{Princeton University Observatory, Princeton, NJ 08544}
\altaffiltext{7}{Institute for Astronomy,
University of Hawaii, 2680 Woodlawn Drive, Honolulu, HI 96822}
\altaffiltext{8}{Department of Astronomy, University of Maryland,
College Park, MD  20742}
\email{bgerke@astron.berkeley.edu}

\begin{abstract}
We use the first 25\% of the DEEP2
Galaxy Redshift Survey spectroscopic data to identify groups and
clusters of galaxies in redshift space.  The data set contains 8370
galaxies with confirmed redshifts in the range $0.7\le z\le 1.4$, over
one square degree on the sky.  
Groups are identified using an algorithm (the Voronoi-Delaunay Method)
that has been shown to accurately reproduce the statistics of groups
in simulated DEEP2-like samples.  
We optimize this algorithm for the DEEP2 survey by applying it
to realistic mock galaxy catalogs and assessing the results
using a stringent set of criteria for measuring group-finding success,
which we develop and describe in detail here.
We find in particular that the group-finder can successfully identify
$\sim78\%$ of real groups and that $\sim79\%$ of the
galaxies that are true members of groups can be identified as 
such.  Conversely, we estimate that $\sim 55\%$ of the groups we find
can be definitively identified with real groups and that $\sim46\%$ of
the galaxies we place into groups are interloper field galaxies. Most
importantly, we find that it is possible to measure 
the distribution of groups in redshift and velocity dispersion,
$n(\sigma,z)$, to an accuracy limited by cosmic variance,  for
dispersions greater than 
$350\kms$.  We anticipate that such measurements will allow strong
constraints to be placed on the equation of state of the dark energy
in the future.  Finally, we present the first DEEP2 group catalog,
which assigns 32\% of the galaxies to 899 distinct groups  
with two or more members, 153 of which have velocity dispersions above
$350\kms$.  We provide locations, redshifts and properties for this
high-dispersion subsample.  This catalog represents the largest
sample to date of 
spectroscopically detected groups at $z\sim 1$.
\end{abstract}


%%%%%%%%%%INTRODUCTION%%%%%%%%%%%%%%%%%%%%%%%%%%%%%%

\section{Introduction}
\label{sec:intro}

Groups and clusters of galaxies are the most massive dynamically
relaxed objects in the universe; as such, they have long been the
subject of intense and fruitful study. 
More than seventy years ago observations of the Coma
cluster gave the first evidence for the existence of dark
matter~\citep{Zwicky}. More recently, studies
of gravitational lensing by clusters have yielded intriguing new
information about the profiles of dark matter halos~\citep{Sand}.
Identifying and studying galaxies in groups and clusters is 
essential to understanding the effects of local 
environment on galaxy formation and evolution \citep[for a review
see][]{BB}. X-ray measurements of the gas mass fraction in clusters
have been used to constrain the mass density parameter $\Omega_M$ and
more recently the equation of state of the dark energy,
$w$~\citep[\emph{e.g.}, ][and references therein]{Allen}. 
Finally, if we can accurately measure the abundance of groups
\emph{and its evolution with redshift}, we can constrain the growth of
large-scale structure, thereby placing significant further constraints
on cosmological parameters \citep{Lilje, ECF, Borgani, HMH, HHM, NMCD}.

A wide array of methods has been used to identify groups and clusters
at moderate redshifts: X-ray emission from hot intracluster
gas~\citep[reviewed by][]{RBN}, cosmic shear due to weak gravitational
lensing~\citep[reviewed by][]{fridge}, searches in optical photometric
data~\citep[\emph{e.g.}, ][]{LCDCS, YG}, the Sunyaev-Zel'dovich (SZ)
effect in the Cosmic Microwave Background~\citep[\emph{e.g.},
][]{LaRoque}, and direct reconstruction of three-dimensional objects
in galaxy redshift surveys~\citep[\emph{e.g.}, ][]{Eke}.  To study the
evolution of the group abundance it is necessary to extend
observations to more distant objects.  However, most of the methods
used for local studies have only limited effectiveness at high
redshift.  The apparent surface brightness of X-ray clusters dims as
$(1+z)^{-4}$, making only the richest clusters visible at high redshift.
The cross-section for gravitational lensing falls rapidly at high
redshifts, 
making weak-lensing detection of distant clusters difficult for
all but the most massive objects.  In 
photometric surveys, the increased depth necessary for high-redshift
studies increases the overall number density of objects, thereby
increasing the problems of foreground and background contamination and
projection effects (though photometric techniques for estimating
redshifts can mitigate these difficulties).  The SZ effect is very
promising, since it is 
entirely independent of redshift, but it also suffers from confusion
limits and projection effects, and in any case a large survey of SZ
clusters is yet to 
be undertaken.  For the time being, then, one of the few methods that
can be applied to large numbers of groups and clusters on similar mass
scales at $z\sim 0$ and $z\sim 1$ is the direct detection of these
structures in the redshift-space distribution of galaxies.

The first sizeable sample of
groups detected in redshift space was presented by~\citet{GH}, who
found 176 groups of three or more galaxies in the CfA galaxy
redshift survey at redshifts $z\la 0.03$. Recently, \citet{Eke}
identified groups  
within the final data release of the Two-degree Field Galaxy Redshift
Survey (2dFGRS).  Their catalog extends to $z\approx 0.25$ and
constitutes the largest currently available catalog of galaxy groups,
containing $\sim 3\times 10^4$ groups with two or more members.  A
comprehensive listing of 
previous studies of local optically selected group samples is also
given by these authors.   Work is currently underway to detect groups
of galaxies in the spectroscopic data of the Sloan Digital Sky
Survey~\citep{Nichol}. Studies of groups detected in redshift space
were extended to intermediate redshifts by~\citet{Carlberg}, who found
more than 200 groups in the CNOC2 redshift survey, with redshifts in
the range $0.1\le z\le 0.55$. Additionally, \citet{Cohen} studied a sample
of 23 density peaks in redshift space, over the redshift range
$0\le z\le 1.25$.  Until now, however, no spectroscopic 
sample has existed with sufficient size, sampling density and redshift 
accuracy to extend redshift-space studies of large numbers of groups
to redshifts $z\ga 0.5$.

The DEEP2 Galaxy Redshift survey \citep{DEEP2, Faber} is the first
large spectroscopic galaxy catalog at 
high redshift, with observations planned for $\sim 5\times 
10^4$ galaxies, most of which will fall in the range $0.7\le z \le
1.4$.  The survey is thus a unique dataset for
studying high-redshift galaxy groups. 
With such a broad range in redshift, we expect to observe
evolution in the properties of galaxies, groups, and clusters within
the DEEP2 sample itself; also, by comparing to local samples from
2dFGRS and the Sloan Digital Sky Survey (SDSS), we expect to observe
evolution between $z\sim 1$ and the present epoch.  DEEP2 is
especially well-suited to studies of groups, since its high redshift
accuracy 
%(spectral resolution $R\approx 5000$, yielding a 
%velocity accuracy $\delta v \approx 25 \kms$, measured from repeated
%observations of some objects) 
allows detailed
studies of their internal kinematics.  Repeated
observations of some DEEP2 galaxies indicate a velocity accuracy
$\delta v \approx 25 \kms$, considerably better than the 2dFGRS value
$\delta v \approx 85 \kms$~\citep{2dFspec} and similar to the $\delta 
v \approx 30 \kms$ attained in the SDSS~\citep{EDR}.  In particular,
it will be possible to estimate the masses of DEEP2 groups from their
velocity dispersions.  We anticipate that, by measuring the evolution
of the group velocity function with redshift, it will be
possible to 
constrain cosmological parameters such as the dark energy density
parameter $\Omega_\Lambda$ and equation of state parameter $w$, as
outlined in \citet{NMCD}.  Before carrying out such studies,
however, it will be essential to develop robust methods for detecting
groups and clusters within DEEP2.

Identifying groups and clusters in redshift space is
well known to be a difficult task.  Most notably, clustering
information is smeared out by redshift space distortions like the
so-called fingers-of-God effect, 
in which galaxies in groups and clusters
appear highly elongated along the line of sight because of
intracluster peculiar motions.   This
intermingles group members with other nearby galaxies and causes
neighboring groups to 
overlap in redshift space.  
A group-finding algorithm that attempts to find all group members will  
thus necessarily be contaminated by interloper field galaxies and will 
necessarily merge some distinct groups together into spurious larger
structures.  Conversely, a group finder that aims to minimize
contamination and over-merging will fragment some larger
clusters into smaller groups.  This trade-off in
group-finding errors is well known \citep[\emph{e.g.}, ][]{NW} and
fundamentally unavoidable.  
It will therefore be essential,
before we begin any program of group finding, to construct a suitable
definition of group-finding success, 
identifying in advance which errors we seek to minimize and
what sort of 
errors we are willing to tolerate.  Ultimately, the chosen definition
of success will depend on the intended scientific purpose of the 
group catalog. A major portion of this paper will be devoted to
defining appropriate measures of group-finding success for the 
DEEP2 survey and optimizing our methods using these criteria.

This paper is organized as follows.  In Section~\ref{sec:DEEP2} we
introduce the DEEP2 sample and discuss the unique opportunities and
difficulties it presents for group finding.  In the same section we
describe the DEEP2 mock galaxy catalogs, which we will use to
calibrate our group-finding methods. Then, in
Section~\ref{sec:optimal}, we describe our criteria for group-finding  
success.  In
Section~\ref{sec:gfchoice}, we give an overview of various
group-finding methods that have been used 
in the literature, and we describe the Voronoi-Delaunay Method (VDM) of
~\citet{Marinoni}, which we will use in this study.   We then proceed
to optimize this method for the DEEP2 sample.  Finally, 
in Section~\ref{sec:catalog}, we 
apply the VDM algorithm to the current DEEP2 observations and present
the first DEEP2 group catalog.

%%%%%%%%%%%%%%%%%%%%%%% SECTION 2 %%%%%%%%%%%%%%%%%%%%%%%%%%%%%

\section{Group finding in the DEEP2 survey}
\label{sec:DEEP2}

\subsection{The DEEP2 sample}

As mentioned, the DEEP2 Galaxy Redshift Survey is the first large
(tens of thousands 
of galaxies) spectroscopic survey of galaxies at high redshifts,
$z\sim 1$.  The goal of 
the survey is to obtain spectra of $\sim5\times 10^4$ objects over
$3.5\;\mathrm{deg}^2$ on the sky to a limiting magnitude of $R_{AB}=24.1$
using the DEIMOS spectrograph on the 
Keck II telescope. Typical redshifts in the survey fall in the range
$0.7 \le z \le 1.4$.   Details of the survey will be described
comprehensively in an upcoming paper by
\citet{Faber}; we summarize the salient information for this study 
here.

The survey consists of four fields on the sky, chosen to lie in zones
of low Galactic dust extinction.  Three-band ($BRI$) photometry has been
obtained for each of these fields using the CFH12K camera on the
Canada-France-Hawaii Telescope (CFHT), as described by~\citet{Coil_b}.
In three of the fields, which each consist of 
three contiguous CFHT pointings covering a strip of 120 arcmin by 30
arcmin, galaxies are selected for
spectroscopy if they pass a simple cut in color-color space.  This
cut reduces the  
fraction of galaxies at redshifts $z<0.7$ to below $10\%$, while
eliminating only $\sim3\%$ of higher-redshift galaxies \citep{Faber}.
A fourth 
field, the extended Groth Survey strip, covers $120$ arcmin $\times
15$ arcmin, and because of a wide variety of complementary
observations underway there, galaxies in this field are targeted for 
spectroscopy regardless of color.  For the sake of uniformity we
neglect the Groth field in the current 
study, though it will be quite useful in future work.  Within each
CFHT pointing, galaxies are selected for spectroscopic observation 
if they can be placed on one of the $\sim40$ DEIMOS slitmasks
covering that pointing.  Within each pointing, slitmasks are tiled in
an overlapping pattern, using an adaptive tiling scheme to increase
the sampling rate in regions of high density on the 
sky, so that the vast majority of galaxies have two
opportunities to be selected for spectroscopy.  Further details of the 
observing scheme can be found in \citet{DGN}.  Overall, roughly
$60\%$ of galaxies that meet our selection criteria are targeted
for DEIMOS observation.

Spectroscopic data from DEIMOS are reduced using an automated
data-reduction pipeline \citep{pipeline}, and redshift identifications
are confirmed visually.  In this paper, we focus 
on galaxies in the three CFHT pointings in which all spectroscopy has
been completed as of this writing.  
The locations of these pointings are given in table~\ref{tab:fields}.
Each pointing has
a width of 48 arcmin in right ascension and 28 arcmin in
declination. These fields have each been fully covered, or very nearly
so, by DEIMOS spectroscopy, with a redshift success rate greater
than $60\%$
for each slitmask and an overall redshift success rate of $\sim70\%$.
These three fields, taken together, comprise a sample of 8785 galaxies  
with confirmed redshifts (8370 with $0.7 \le z \le 1.4$), with a
median redshift of $z=0.912$. 
This sample represents the largest sample ever used for group-finding
in redshift surveys of distant ($z\ga 0.25$) galaxies, being more than
twice as large as the CNOC2 sample of \citet{Carlberg}, and the first
such sample at $z\sim 1$. 

\begin{deluxetable}{ccccc}
\tabletypesize{\small}
\tablewidth{0pt}
\tablecaption{Locations and observational status of the DEEP2 pointings
considered in this paper.\label{tab:fields} }
\tablehead{
\colhead{Pointing}&\colhead{}  &\colhead{}    & \colhead{Redshift}\\ 
\colhead{name}\tablenotemark{a} & \colhead{RA}\tablenotemark{b} 
& \colhead{dec}\tablenotemark{b} & \colhead{success}\tablenotemark{c}
}
\startdata
22 & 16 51 30 & +34 55 02 & 0.72 \\
32 & 23 33 03 & +00 08 00 & 0.71 \\
42 & 02 30 00 & +00 35 00 & 0.70 
\enddata
\tablenotetext{a}{The pointings are named according to a
convention in which, for example, pointing 32 refers to the second
CFHT photometric pointing in the third DEEP2 field.}
\tablenotetext{b}{Positions of the pointing centers 
sky are given in J2000 sexagesimal coordinates.}
\tablenotetext{c}{Fraction of spectroscopic targets for which a
definite redshift could be measured.}
\end{deluxetable}

\subsection{The DEEP2 mock catalogs}
\label{sec:mocks}

Both to assess the impact of selection effects and to test and
calibrate our group-finding, it will be necessary to study the
properties of groups in realistic mock galaxy catalogs.  For this
purpose we will 
use the mock catalogs developed by~\citet{YW}.  These catalogs are
produced by assigning ``galaxies'' to N-body simulations according to
the prescriptions of the popular ``halo model'' for large-scale
structure formation \citep[\emph{e.g.}, ][]{Seljak, PS}.   This model
assumes  
that all galaxies form within virialized dark matter halos.  The mean 
number of galaxies above some luminosity $L_{\mathrm{cut}}$
in a halo of mass $M$ is then given by the Halo  
Occupation Distribution $N(M)$ \citep{Berlind, MH}, while the
luminosities of galaxies in 
the halo obey a Conditional Luminosity Function, 
$\Phi(L|M)$ \citep{YMvdB}, which 
is allowed to evolve with redshift in keeping with observations.  These
functions can be varied to produce mock galaxy catalogs that match the
observed DEEP2 redshift distribution and clustering statistics, as
measured by~\citet{Coil}.  In the DEEP2 mock catalogs used here,
the ``galaxies'' populating a given host halo are assigned positions and
velocities as follows: the brightest galaxy in a halo is placed at the
halo's center of mass, and all other galaxies are assigned to random
dark matter particles within the halo.

For the purposes of this work, we will use the most recent version of
the mock catalogs produced
using simulation 4 from Table 1 of~\citet{YW}; for further details
about the creation of the DEEP2 mock galaxy catalogs, the reader is
referred to that paper. 
Here we merely note in summary that the catalogs comprise twelve nearly
independent mock DEEP2 fields with the same geometry as the
three high-redshift DEEP2 fields, extending over a redshift range
$0.6\la z\la 1.6$.  They have been constructed by 
populating N-body simulations computed in a flat $\Lambda$CDM
cosmology with density parameter $\Omega_M$=0.3, fluctuation amplitude
$\sigma_8$=0.9, spectral index $n$=0.95, and dimensionless Hubble
parameter $h=H_0/100\kms=0.7$.  
The evolution of large-scale structure
with redshift is included in the mocks by stacking different
time slices from the N-body simulations along the line of sight.
The simulations resolve dark matter halos down to
masses around $8\times 10^{10} M_\odot h^{-1}$, sufficiently low to
encompass all galaxies above $L_{\mathrm{cut}} = 0.1L_*$.  This
luminosity cut, in turn, is sufficently low to be below the DEEP2
magnitude cut for the redshift range of interest here, $z\ge 0.7$.  
We have tested our group-finding methods on mock catalogs created
using different halo model parameters, and we find that the results
presented in this paper are  
acceptably robust to such changes (\emph{i.e.}, their effects on the
reconstructed group catalog are generally smaller than the cosmic
variance).

In order to study the impact of galaxy selection effects on our group
sample, we produce four distinct subsamples from the mock catalogs. 
The \emph{volume-limited} sample contains all galaxies down to a
limiting magnitude $L_{min}=0.1L_*$ (it is important to
note that this catalog is not ``volume-limited'' in the traditional sense,
since $L_*$---and hence $L_{min}$---varies with redshift in the mock
catalogs).  The \emph{magnitude-limited} sample has had the DEEP2
magnitude limit of $R_{AB}<24.1$ applied, cutting out the faint
galaxies in the volume-limited sample in a distance-dependent way (the
mock catalogs do not contain color information, so no color cut is
applied; we simply take the DEEP2 color criteria to be equivalent to
the redshift limit $z>0.7$).
The \emph{masked} sample is the result of applying the DEEP2
``mask-making'' algorithm (see \citet{DGN} and \citet{Faber} for
details), which schedules galaxies for slitmask 
spectroscopy, to the magnitude-limited sample.  Because the amount of
space on DEIMOS slitmasks is finite, and because neighboring slits'
spectra may not overlap, only $\sim60\%$ of suitable target galaxies
can be scheduled for observation. 

Finally, the \emph{mock DEEP2} sample
simulates the effects of redshift failures within the observed DEEP2
sample.  Currently, approximately $30\%$ of observed DEEP2 galaxies
cannot be assigned a firm redshift, in large part because of the
presence 
of galaxies at $z\ga 1.5$, for which no strong spectral features fall
in the DEEP2 wavelength range, but also because of poor observing
conditions, low signal-to-noise ratio, or instrumental effects.  The
redshift 
success rate also has some magnitude dependence for faint galaxies,
dropping by $\sim15\%$ between $R=22.6$ and $R=24.1$.  These effects
are fully taken into account in the mock DEEP2 sample.  Since this
sample is the most similar to the actual DEEP2 redshift catalog, we
will use it to test and calibrate our group-finding algorithm; we
will use the other three samples to study various selection effects.

\subsection{Difficulties for group finding in deep redshift surveys}
\label{sec:difficulties}

Identifying an unbiased sample of groups and
clusters of galaxies in redshift space is notoriously difficult.  
As mentioned in Section~\ref{sec:intro}, the most obvious and
well-known complication is redshift-space distortions: the 
orbital motions of galaxies in virialized groups cause the observed
group members to appear spread out along the line of sight (the
fingers-of-God effect), while coherent infall of
outside galaxies into existing groups and clusters reduces their
separation from group centers in the redshift direction
(the Kaiser effect). 
Both of these effects confuse group membership by intermingling
group members with other nearby galaxies.  Since it is
impossible to separate the peculiar velocity field from the Hubble
flow without an absolute distance measure, this confusion can never be
fully overcome, and it will be a significant source of error in any
group-finding program in redshift space.  A second complication arises
from incomplete sampling of the galaxy population.  No modern galaxy
redshift survey can succeed in measuring a redshift for every target
galaxy, and it has been shown \citep{SzSz} that an incomplete galaxy
sampling rate always leads to errors in the reconstructed catalog of
groups and clusters---even without redshift-space distortions. 

In addition, surveys conducted at high redshift
and over a broad redshift range present their own impediments
to group finding.  The first is simple: distant galaxies appear
fainter than nearby galaxies.  For example, the DEEP2 $R_{AB}=24.1$ 
magnitude limit 
means that the faintest DEEP2 galaxies at $z\sim1$ have luminosities
near $L_*$ \citep{Willmer}.  We are thus probing only relatively rare,
luminous galaxies,  
so only a small fraction of a given group's members will meet our
selection criteria.  Moreover, galaxies selected with the same
criteria will correspond to different samples at different
redshifts. Selection in the $R$ band (as is done for DEEP2) 
corresponds to a rest-frame $B$ band selection at $z = 0.7$ and a
rest-frame $U$ band selection at $z\ga 1.1$, meaning
that red, early-type galaxies will drop below the limiting magnitude at
lower redshifts than blue, star-forming galaxies.  Since
blue galaxies are observed to be less strongly clustered
than red galaxies in DEEP2 \citep{Coil} and locally \citep[\emph{e.g.},
][]{Madgwick_a}, we  
expect that the density contrast between group members and isolated
galaxies will be weaker for the DEEP2 sample than it would be for
a sample selected in rest-frame $I$, for example.  Finally, the very
evolution of large-scale structure with 
redshift that one wishes to probe will pose a problem, since the mass
function of dark matter halos will be shifted to lower masses at high
redshift, leading to smaller groups and clusters.  

\begin{figure}
\centering	
\epsscale{1.1}
\plotone{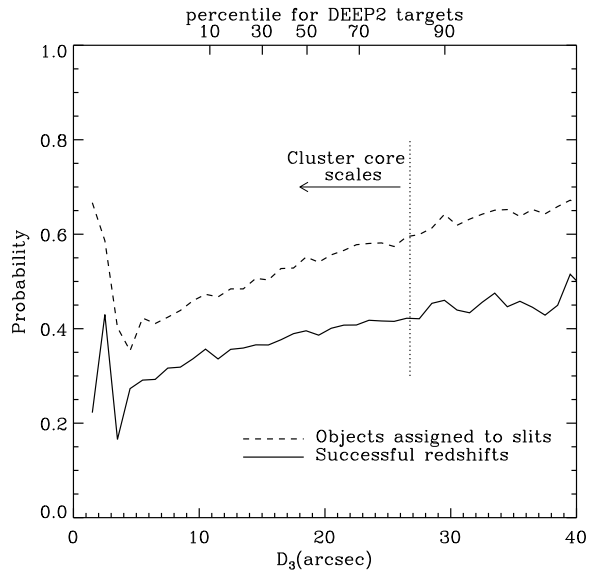}
\caption{Rates of spectroscopic observation and redshift success as a 
function of local density of DEEP2 target galaxies on the sky, as
measured by the distance $D_3$ of a galaxy from its third-nearest
neighbor.  The 
dashed line shows the probability that a galaxy meeting the DEEP2
targeting criteria is scheduled for spectroscopic observation, as a
function of 
$D_3$.  The solid line shows the probability that an observed galaxy
yields a successful redshift,
multiplied by the dashed line, to give the total probability that a 
potential DEEP2 target has its redshift measured, as a function of
$D_3$.   The top axis shows 
the percentage of DEEP2 galaxies that have $D_3$ less than the
indicated value, and the dotted line shows the scale of a typical
cluster core ($300$ kpc) at $z=1$.  Clearly, the 
probability of observation is reduced in regions of high local
density, although local density appears to have little further effect
on redshift success.  The sharp increase in the ratios at very low
$D_3$ arises because extremely close pairs of galaxies may be observed
together on a single slit.} 
\label{fig:neigh}
\end{figure}

A further, more complicated problem is posed by the realities of
multi-object spectroscopy.  Because of the physical limitations of
slitmask or fiber-optic spectrographs, it is difficult to observe
all galaxies in densely clustered regions.  In DEEP2, for 
example, the minimum DEIMOS slit length is three
arcseconds (approximately $20$ kpc at $z\sim1$); objects closer than
this on the sky cannot be observed on 
the same slitmask (except in the special case of \emph{very} close
and appropriately aligned neighbors, which can both be observed on a
single slit).  This problem is mitigated somewhat by the adaptive
scheme for tiling the DEEP2 CFHT imaging with slitmasks, which gives
nearly every target at least two chances to be observed; nevertheless,
slit 
collisions cause us to be biased against observing objects that are
strongly clustered on the sky. 
Moreover, the quality of DEIMOS spectra is degraded somewhat for short
slit lengths, due to the difficulty of subtracting night-sky emission
for such slits,
so we might expect a lower redshift success rate for clustered
objects.

Figure~\ref{fig:neigh} shows the probabilities of observation and 
redshift success as functions of the distance to an object's
third-nearest  neighbor on the sky. (We have chosen the
\emph{third}-nearest neighbor distance because this is a less noisy
measure of local density than the simple nearest-neighbor distance.)
Clearly we are less likely to 
observe galaxies in dense regions on the sky, though this effect is
relatively weak, and local density
appears to have little effect on the redshift success rate.  Moreover,
as shown in the figure, the vast majority of DEEP2 targets have
neighbors on the sky at distance scales smaller than a typical cluster
core radius ($\sim 300$ kpc).  Since we expect a much smaller
percentage of galaxies to 
actually reside in cluster cores, we conclude that a given
galaxy's close neighbors are frequently in the foreground or
background.  Hence, although we clearly undersample galaxies in dense
regions \emph{on the sky}, we are not necessarily undersampling
galaxies in dense regions in three-space.  Nevertheless, 
all of the effects discussed in this section, taken together, mean
that \emph{nearly all DEEP2 groups will have fewer than ten members}
(see Table~\ref{tab:groups}).

\begin{figure}
\centering	
\epsfig{width=2.75in, angle=90,file=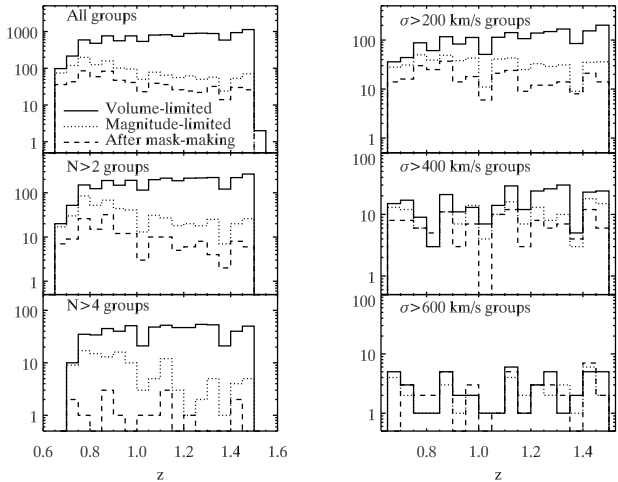}
\caption{Distribution of group redshifts in a
  mock DEEP2 field ($120\times 30$ arcmin).  \emph{Left}: The upper
  panel shows redshift distributions for all groups 
  that enter a given catalog with richness $N\ge 2$.   The   
  solid line shows the distribution for the volume-limited catalog,
  the dotted line shows the distribution for the magnitude-limited
  catalog, and the dashed line shows the distribution for the masked
  catalog (see Section~\ref{sec:mocks} for the definitions of these mock
  samples).   The apparent decrease in group abundances at low
redshifts arises is intended to mimic the DEEP2 photometric selection
criteria.  The middle panel
  shows the distributions for groups entering each catalog with
  richness  $N>2$, and the bottom panel shows the distributions for
  groups with $N>4$. 
  \emph{Right}:  Redshift distributions for the same three catalogs,
  for groups with velocity 
  dispersion above some threshold $\sigma_c$.  From top to bottom, the
  panels represent $\sigma_c=200$, $400$ and $600\;\kms$.  Note that,
  when groups are selected by velocity dispersion, the discrepancy
  between the three catalogs decreases as $\sigma_c$ increases,
  whereas the discrepancy increases with richness.}
\label{fig:dropoff}
\end{figure}

The galaxies in each group will thus represent a very sparse, discrete
sampling of the membership of each group.  It is well known
that large errors can result when the moments of a distribution are
estimated from a sparse sample.  In particular,
computing velocity dispersions with the usual formula for standard
deviation, $\sigma^2=\langle v^2 \rangle - \langle v \rangle^2$, will
be an unreliable method for such small groups.  \citet{BFG} have
studied this issue in the 
context of galaxy clusters.  They assess a number of alternative
dispersion estimators and determine the most accurate ones for
different ranges in group richness.  For the richness range of
interest here, $N\sim5$, they find the most robust method to be
the so called ``gapper'' estimator, which measures velocity dispersion
using the velocity gaps in a sample according to the formula
\begin{equation}
\sigma_G = \frac{\sqrt{\pi}}{N(N-1)}\sum_{i=1}^{N-1}i(N-i)(v_{i+1}-v_i), 
\label{eqn:gapper}
\end{equation}
where the line-of-sight velocities $v_i$ have been sorted into
ascending order. 
Since we expect this estimator to be more accurate than the standard
deviation for our purposes, we will measure velocity dispersions as
$\sigma = \sigma_G$ throughout this paper.  Furthermore, in this paper
we shall always compute velocity dispersions using the \emph{galaxies}
in a given sample.  Correcting these values to reflect the
velocity dispersions of dark matter halos 
will ultimately be necessary for comparison with predictions, but we
focus here on measureable quantities and defer this (theoretical) issue
to future work.

The effects of the DEEP2 target selection criteria can be seen in
Figure ~\ref{fig:dropoff}.  The upper left-hand panel
shows the redshift distribution of 
groups in a single mock DEEP2 pointing, drawn at random from the
mock catalogs, for the volume-limited, magnitude-limited, and masked
samples.  Here a \emph{group} is defined to be the set of all galaxies
in a given sample
that occupy a common dark matter halo, and a group's redshift is given
by the median redshift of its member galaxies.  The remaining panels
show subsets of these three group 
catalogs, containing groups above a given threshold in richness $N$ or 
line-of-sight velocity dispersion $\sigma$.  It is worth noting
briefly that,  
in some redshift bins, the masked sample has \emph{more} groups than
the magnitude-limited sample from which it is drawn.  This effect
is easy to understand: it occurs when group members are discarded,
moving the median redshifts 
of some groups from one bin to another.  The important point, however,
is that when the $\sigma$ threshold is increased, the discrepancies
become smaller between the volume-limited, magnitude-limited and
masked samples.  On the other hand, these discrepancies increase when
the richness threshold is increased: we note in particular the sharp
drop-off in groups with $N>4$ between the magnitude-limited and masked
sample.  Evidently groups selected according to observed richness
constitute a significantly biased sample, whereas groups selected by
observed velocity dispersion can provide a more accurate
representation of the full underlying sample. 

\begin{figure}
\centering
\epsfig{width=3in,file=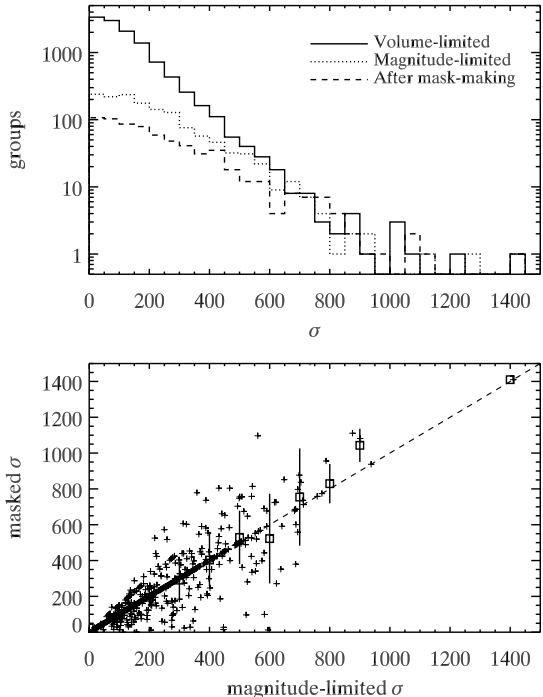}
\caption{
Effects of DEEP2 selection on group velocity dispersion
$\sigma$. 
The upper panel shows the $n(\sigma)$ distribution for groups in each
of the volume-limited, magnitude-limited and masked samples in a
single mock DEEP2 field.  Note that, as suggested by
figure~\ref{fig:dropoff}, the three distributions are similar at high
velocity dispersions.  The lower panel shows how
individual groups' velocity dispersions change when the DEEP2
slitmask-making algorithm is applied. Crosses show the dispersions of
individual groups computed from the galaxies present before and after
mask-making; open squares 
and error bars show the mean and standard deviation of the masked
$\sigma$ value in 
bins of $100\kms$ in magnitude-limited $\sigma$ value.  The dashed
line is the line of equality for the pre- and post-maskmaking velocity 
dispersions.  A majority ($57\%$) of the groups plotted fall
exactly on this line.}
\label{fig:group_shrink}
\end{figure}

This result is not surprising.    Velocity 
dispersion is known to scale with halo mass roughly as $\sigma \propto 
M^{1/3}$ \citep{BN}, and richness should also scale with $M$.
In a magnitude-limited sample, measured group richnesses will be affected
by the flux limit so that more distant groups will have fewer observed
members: for example, a group observed to have three
members at $z=1.3$ will actually contain significantly
more galaxies than a three-member group observed at $z=0.7$.
However, sufficiently massive 
(\emph{i.e.}, high-dispersion and -richness) groups are nearly
certain to enter the observed catalog with more than one member at all
redshifts---and hence to be identifiable as groups.  
Selection effects that 
reduce the number of galaxies observed in each group 
will introduce a scatter in the measured velocity dispersions of
individual groups.  But above some appropriate critical dispersion,
$\sigma_c$ we expect the observed \emph{distribution} of group
velocity dispersions, $n(\sigma)$, to resemble the true one.  
In Figure~\ref{fig:group_shrink}, we see that this expectation is
borne out in the DEEP2 mock catalogs. Although a significant 
scatter exists in measured group velocity dispersions, the agreement
between the $n(\sigma)$ distributions for the three mock catalogs
improves with increasing $\sigma$.  For these reasons, we expect 
that it will be possible to identify a robust sample of DEEP2 groups 
whose $n(\sigma)$ distribution is not strongly biased by observational
effects.

%%%%%%%%%%%%%%%%%%%%%%%%%%%%%%%% SECTION 3 %%%%%%%%%%%%%%%%%%%%%%

\section{Defining the optimal group catalog}
\label{sec:optimal}

Because we expect any group-finding algorithm to be prone to many
different types of error, it is crucial that we define carefully our
tolerance for various errors and craft a specific definition of
group-finding ``success.''  To begin with, we must
establish what we mean when we speak of a galaxy group. 
As already noted briefly in Section~\ref{sec:mocks}, in the spirit of
the halo model, 
we define galaxy groups in terms of dark matter halos.  We define a
\emph{parent halo} to be a single, virialized halo that contributes
one or more galaxies to our sample; the contributed galaxies 
we call the halo's \emph{daughter galaxies}.  A \emph{group}
is then defined to be a set of (two or more) galaxies that comprises
the daughter galaxies of a single parent halo.  \emph{Field galaxies}
are those galaxies that constitute the lone daughters of their
respective parent halos. These definitions are convenient because
cosmological tests 
based on cluster abundance are 
in reality concerned with the abundance of virialized dark matter
halos; we wish to infer the presence of such objects from the
clustering of galaxies. 
In applying this definition we consider to be separate groups those
halos that are 
not virialized with respect to each other in a common potential well,
but we make no distinction between subhalos within a larger, common
virialized halo.  It will also be necessary in what follows to
differentiate between \emph{real groups}---those sets of galaxies that
actually share the same underlying dark matter halo---and
\emph{reconstructed groups}---the sets of galaxies identified as
groups by the group finder.

The ideal reconstructed group catalog would be one in which (\emph{i}) all
galaxies that belong to real groups are identified as group members,
(\emph{ii}) no field galaxies are 
misidentified as group members, (\emph{iii}) all
reconstructed groups are associated with real, virialized 
dark-matter halos, (\emph{iv}) all real
groups are identified as distinct objects, 
and (\emph{v}) these objects contain all of their daughter galaxies
and no others.  As discussed in Section~\ref{sec:difficulties},
however, such a catalog is impossible to achieve because of
redshift-space distortions and incomplete sampling of the galaxy
population.  Nevertheless, this ideal will
be useful as a means of assessing the veracity of our group catalog.
It is thus important to define a vocabulary with which to compare our
group catalog to the ideal one.  We shall make frequent use
of the following definitions:  a group catalog's \emph{galaxy-success
rate} $S_{Gal}$ is the fraction of 
galaxies belonging to real groups that are identified as members of
reconstructed groups.   \emph{Interlopers} are field galaxies that are
misidentified as group members in the reconstructed catalog, and the
\emph{interloper fraction} $f_I$ of a group catalog is the fraction of
reconstructed group 
members that are interlopers.  The \emph{completeness} $C$ of a group
catalog is the fraction of real groups that are successfully
identified in the reconstructed catalog (we shall define what it means
to be ``successfully identified'' shortly); conversely, the
\emph{purity} $P$ is the fraction of reconstructed groups
that correspond 
to real groups.  \emph{Fragmentation} occurs when a real group is
identified as several smaller groups in the reconstructed catalog, and
\emph{over-merging} occurs when two or more real groups are identified
as a single reconstructed object.

Since a perfect group catalog is impossible to achieve, we shall
focus our efforts on reproducing certain selected group properties as
accurately as possible.   There are many properties we could choose to
reproduce for different scientific purposes; each choice has
advantages and drawbacks.  We could, for example, choose to maximize
$S_{Gal}$, thus ensuring that our group catalog contains all
galaxies that belong to real groups.  Such a sample would likely have
a high interloper fraction and much over-merging, however (for
example, the easiest way to ensure $S_{Gal}=1$ would be simply to 
place all galaxies in the sample into a single group).
Conversely, a group 
catalog that minimizes the interloper fraction would
likely be highly incomplete, sucessfully finding only the cores of the
largest groups.
Such catalogs might be useful for studies of the properties of galaxies
in groups, but they are unlikely to be of much use for studying
cosmology or large-scale structure.

A different approach is to gauge success on a group-by-group basis and
attempt to maximize completeness, purity, or both.  To do
this, we must develop a quantitative measure of our success at
reconstructing individual groups; we will use the concept
of the Largest Group Fraction (LGF) 
\citep[\emph{cf.}][and references therein]{Marinoni}.
To compute the LGF for a given real group $G$, we first 
find the reconstructed group $G^\prime$ that contains a plurality of
the galaxies in $G$ (the fact that this is not necessarily
unique does not 
concern us, since we will eventually require a majority for a
successful reconstruction). The group $G^\prime$ we call the
\emph{Largest Associated Group} (LAG) of $G$.   The LGF
$\mathcal{L}_G$ of group $G$ is then defined as   
\begin{equation}
\label{eqn:lgf}
\mathcal{L}_G=\frac{N(G\cap G^\prime)}{N(G)},
\end{equation}
where the notation $N(A)$ denotes the number of galaxies in the set
$A$.  That is, the LGF is the fraction of group $G$ that is contained
in its LAG $G^\prime$.  The LGF of a reconstructed group is defined
similarly, but with $G$ 
being drawn from the reconstructed catalog and its LAG $G^\prime$
being drawn from the real catalog.  It should be mentioned here that
$\mathcal{L}_G$ can only be measured for groups in mock 
catalogs, where we know the real group memberships.  In all further 
discussion of tests involving the LGF, it should be assumed that these
tests take place in mock catalogs.  

\begin{figure}
\centering
\epsfig{file=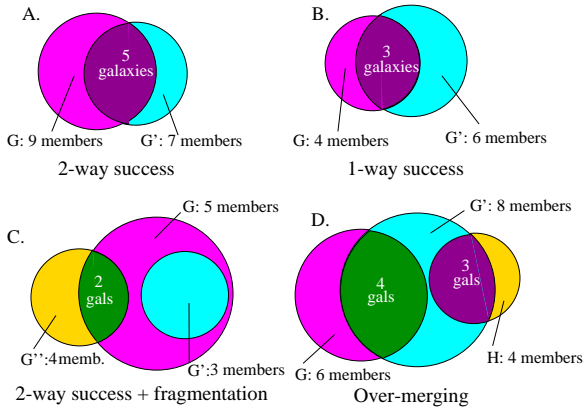, width=3in}
\caption{Schematic depiction of various success and failure modes
for group-finding under the criteria discussed in the text.
Diagrams show hypothetical comparisons between real groups (unprimed)
and the groups found by a group finder (primed).  \emph{(A)} A fully
two-way successful reconstruction, in which both the real and found
group have $\mathcal{L}_G > 0.5$.  \emph{(B)} A one-way success, 
in  which the real group has $\mathcal{L}_G > 0.5$, but 
half of the found group is made up of interlopers.  \emph{(C)} Both a
two-way success ($G$ and $G^\prime$) and a failure due to fragmentation
($G^{\prime\prime}$).  \emph{(D)} An example of overmerging, in which
$G$ and $H$ are both one-way successful real groups, but $G^\prime$
combines their members into a single group.}
\label{fig:successfig}
\end{figure}

The LGF statistic allows us to define an unambiguous set of
group-finding success measures.  In essence, we declare a successful
detection if a group's $\mathcal{L}_G$ is greater than some fraction
$f$. For this 
definition to be unique, we must have $f \geq 0.5$, with higher values
of $f$ implying a more stringent definition of success.  For the
remainder of this work, we will set $f$ to the minimal value of $0.5$,
since, as we shall see, this definition of success is already quite
strict.  However, 
simply requiring a group to have  $\mathcal{L}_G>f$ is insufficient: a
real group could meet this criterion but still been merged
into a larger object by the group-finder, and a reconstructed group
$G^\prime$ 
could have $\mathcal{L}_{G^\prime}>f$ if it is a fragment of a larger real
group.  For this reason, it will be
important to differentiate between one-way matches, in which a group
simply has LGF above $f$, and two-way matches, in which a group $G$
and its LAG $G^\prime$ satisfy $\mathcal{L}_G,
\mathcal{L}_{G^\prime}>f$, \emph{and} 
$G$ is also the LAG of $G^\prime$ (See Figure~\ref{fig:successfig}
for a schematic depiction of these success measures).  Hence we shall
differentiate between 
one-way purity $P_1$, the fraction of reconstructed groups
with $\mathcal{L}_G > f$, and two-way purity $P_2$, the fraction of
reconstructed groups which are two-way matches with some real group. 
Similarly, for real groups, we define one-way completeness $C_1$ and two-way
completeness $C_2$.  Comparing these
statistics can give some indication of systematic errors in the group
catalog.  A real group that
is a one-way success but \emph{not} a two-way success has likely been
overmerged by the group finder; therefore if $C_1$ is much larger than
$C_2$ we 
expect that our catalog has been highly over-merged.  Similarly,
if $P_1$ is significantly greater than $P_2$, we expect that our
catalog is highly fragmented.

Our definition of success has another potential problem, however: 
it requires, minimally, only that we reconstruct half of each group.
Thus, a ``successful'' search strategy could seek only the most
tightly clustered sets of galaxies and detect only the
cores of groups and clusters.  Such a group catalog would likely be
of high purity, with few interlopers; it could be 
useful for identifying groups for follow-up observation in X-ray or
Sunyaev-Zeldovich surveys.  But it would likely have low
completeness, and it probably would not accurately
reproduce group properties like richness, physical size, or velocity
dispersion, making estimates of cluster mass impossible with
spectroscopic data alone. 

In order to mitigate such difficulties,
we must also monitor our success in reproducing group properties.
In part, this means we should attempt to accurately measure
properties like 
the velocity dispersion of successfully reconstructed groups on a
group-by-group basis. However, since errors in individual group
detections are inevitable 
with any group finder, we must also determine whether these
errors bias the overall distribution of group properties in our
catalog.  Ultimately it is these statistical \emph{distributions} we
will want to reproduce as accurately as possible.  For 
example, if we wish to study the abundance of groups as a function of
redshift, $n(z)$, we must take care to ensure that spurious group
detections and undetected real groups do not skew this distribution.  

Clearly, then, there are many different possible means by which we
could gauge our success at group finding.  As we have said, our chosen
measure of 
success will depend strongly on the ultimate scientific purposes of
our group catalog.   In our case, among other uses, we
envision using the DEEP2 group catalog to constrain cosmological
parameters. \citet{NMCD} 
have shown that DEEP2 groups can be used for this purpose if their
abundance is measured accurately as a bivariate distribution in
velocity dispersion and redshift, $n(\sigma,z)$.  It has
also been shown \citep{Marinoni} that the Voronoi-Delaunay group-finding
algorithm can successfully reconstruct this distribution (down to some
limiting 
velocity dispersion $\sigma_c$); hence we will seek in this study to 
maximize the accuracy of our reconstructed $n(\sigma,z)$ above some
$\sigma_c$.  Of course, 
it would be possible in principle to reproduce this distribution by
chance with a low-purity, low completeness catalog.  Therefore, we will
simultaneously strive to maximize the completeness and purity
parameters, while also taking care to keep  $C_1\approx C_2$ and
$P_1\approx P_2$ to guard against fragmentation and over-merging. 
Indeed, it is always important to monitor these statistics in order
to ensure that our reconstructed group catalog corresponds reasonably
well with reality.   We do not actively monitor the $S_{Gal}$ or
$f_I$ parameters when optimizing our group finder, but we anticipate
that a catalog that meets our 
success criteria will also be of reasonably high quality by these
measures as well (this is borne out in Section~\ref{sec:optimize}).

In concluding this section, it is important to note that when we speak
of group velocity dispersions or redshifts in this paper, we are
talking only about the properties as computed from the \emph{observed
group members}.  Although we will ultimately be interested in the
properties of dark matter halos (which can be predicted
theoretically and used to constrain cosmology), these cannot be
measured directly, even in principle. Even with a 
completely error-free group catalog, a theoretical correction would
have to be applied to account for the effects of discreteness. 
Thus we will be interested in reconstructing $n(\sigma,z)$ \emph{as
computed using observed galaxies only}.  We make no attempt to
reconstruct the distribution as computed for the dark matter or using
unobserved galaxies, and we leave computation of theoretical
correction factors to future work.

%%%%%%%%%%%%%%%%%%%%%%%%%  SECTION 4 %%%%%%%%%%%%%%%%%%%%%%%%%%%

\section{Choosing and optimizing the group-finder}
\label{sec:gfchoice}

Several different techniques have been developed to
find groups in spectroscopic redshift samples.  We review them
briefly here, as a means of introducing the main issues that will
concern us in selecting a group-finding algorithm. 

\subsection{A brief history of group finding}
\label{sec:groupfinders}

\citet{HG} presented a simple early method for identifying groups and
clusters in the Center for Astrophysics (CfA) redshift survey by
looking for nearby neighbor galaxies around each galaxy. 
Commonly known as the \emph{friends of friends} or \emph{percolation}
method, this technique, in its simplest form, defines a linking length
$b$ and links every galaxy to those 
neighboring galaxies a distance $b$ or less away (``friends'').  This
procedure produces complexes of galaxies linked together via their
neigbors (``friends of friends''); these complexes are identified as
groups and clusters.  Versions of this algorithm have been widely used
to identify groups in local redshift surveys---most recently by
\citet{Eke} in 2dFGRS---and percolation techniques have also long been
used to identify virialized dark-matter halos within N-body
simulations.  The 
percolation algorithm is intuitively attractive because
it identifies those regions with an overdensity $\delta \ge
(2\pi b^3/3)^{-1}$ compared to the background density.  The
overdensity $\delta_v$ of virialized objects can be 
readily computed using the well-known spherical collapse model,
yielding an appropriate linking length of $b=0.2\langle\nu\rangle^{-1/3}$ for
identifying virialized objects in an 
Einstein-de Sitter universe \citep{DEFW}, where $\langle\nu\rangle$ is the
mean spatial number density of galaxies (this linking length is
somewhat smaller for a $\Lambda$CDM model, a point
which has frequently been ignored in the literature). Hence, the
percolation algorithm is a natural method for identifying virialized
structures in the absence of redshift-space distortions.

Unfortunately, working in redshift space can cause
serious problems for this algorithm.  The fingers-of-God
effect requires that we stretch the linking volume into an ellipsoid
or cylinder along the line of sight, which increases the possibility
of spurious links.
Because the percolation method considers each galaxy equally while
creating links, then places \emph{all} linked 
galaxies into a given group or cluster, such false links can lead to
catastrophic failures, in which the group finder ``hops'' between
several nearby groups, merging them together into a single, falsely
detected 
massive cluster.  On the other hand, shrinking the linking volume to
avoid this problem
increases the chances that a given structure will be fragmented into
several smaller structures by the group finder or missed entirely.
These problems have been studied in detail by \citet{NW} and more
recently by \citet{Frederic}.

To combat such difficulties, various other group-finding methods have been
developed.  \citet{Tullya, Tullyb} used the so-called ``hierarchical''
group-finding scheme, originally introduced by \citet{Materne}, to find
nearby groups.  The hierarchical grouping
procedure used is computationally interesting, but in the
context of the current model of structure formation it seems to lack
theoretical motivation.  More recently, the SDSS team has 
introduced a group-finding algorithm called C4 \citep{Nichol}, which
searches for clustered galaxies in a seven-dimensional space,
including the usual three redshift-space dimensions and four
photometric colors, on the principle that galaxy clusters should
contain a population of galaxies with similar observed colors.
\citet{Kepner} 
introduced a  three-dimensional ``adaptive matched filter'' algorithm
which identifies clusters by 
adding ``halos'' to a synthesized background mass density and
computing the maximum-likelihood mass density.  \citet{WK} found that
this algorithm is extremely successful at identifying clusters in
spectroscopic redshift surveys, and recently, \citet{Yang} have
introduced a group-finder that combines elements of the matched filter
and percolation algorithms.  Finally, \citet{Marinoni} developed a
group-finding algorithm---the Voronoi-Delaunay Method (VDM)---that
makes use of the Voronoi partition and Delaunay triangulation of a
galaxy redshift survey to identify high-density regions.  By
performing a targeted, adaptive search in these regions, the VDM
avoids many of the  pitfalls of simple percolation methods;  
we will use a version of it in this study.
We note in passing, however, that the matched-filter algorithm is also
attractive for DEEP2, and we plan to explore its usefulness in future
studies.

\subsection{The Voronoi-Delaunay method}
\label{sec:VDM}

\begin{figure}
\centering
\epsfig{width=3.4in,file=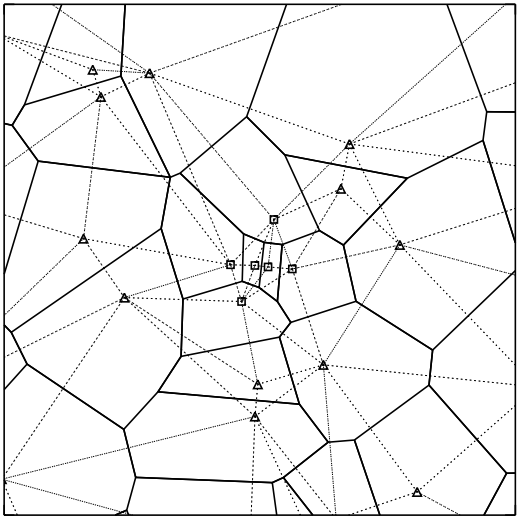}
\caption{The two-dimensional Voronoi partition (Dirichlet tesselation)
and Delaunay mesh for an array of points.  The points consist of a
randomly generated uniform background (triangles) and a small, tightly 
clustered group of points (squares) that roughly approximate a galaxy 
group.  Dotted lines show the Delaunay mesh, which connects
each point to its nearest neighbors.  The solid lines delineate the    
edges of the Voronoi polygons---the perpendicular bisectors of the
Delaunay links.  Note that each polygon contains only one point, and
that the typical Voronoi cell is smaller for the grouped points than
for the background points.}
\label{fig:voronoi}
\end{figure}

\citet{Marinoni} showed that the VDM successfully reproduces the
distribution of groups in velocity dispersion in a DEEP2-like sample
down to some minimum 
dispersion $\sigma_c$. The algorithm makes use of the
three-dimensional Voronoi partition of the galaxy redshift catalog,
which tiles space with a set of unique polyhedral sub-volumes, each of
which 
contains exactly one galaxy and all points closer to that galaxy than
to any other.    The Voronoi partition naturally provides
information about the clustering properties of galaxies, since
galaxies with many neighbors will have small Voronoi volumes, while
relatively isolated galaxies will have large Voronoi volumes.  The
algorithm also makes use of the clustering information encoded in the
Delaunay mesh, which is a complex of line segments linking neighboring
galaxies.  Mathematically speaking, the Delaunay mesh is the
geometrical dual of the Voronoi partition; the faces of the
Voronoi cells are the perpendicular bisectors of the lines in the
Delaunay mesh.  A two-dimensional visual representation of the Voronoi
partition and 
Delaunay mesh is shown in Figure~\ref{fig:voronoi}. 

The VDM group-finding algorithm proceeds iteratively through the
galaxy catalog in three phases as follows.
In Phase I, all galaxies that have not yet been assigned to groups 
are sorted in ascending order of Voronoi volume.  This is mainly a
time-saving step, because it allows us to begin our group search with
those galaxies in dense regions.  Then, for the first galaxy in this
sorted list (the \emph{seed} galaxy), we define a relatively small
cylinder of radius 
$\mathcal{R}_{\mathrm{min}}$ and length $2\mathcal{L}_{\mathrm{min}}$,
oriented with its axis along the redshift direction.  The dimensions
of this and other search cylinders are computed using comoving
coordinates.\footnote{One might naively expect to use
\emph{physical} coordinates to find virialized objects like clusters,
but because the background density scales as $\rho_b\propto(1+z)^3$,
dark matter halos of a given mass have virial radii that scale
roughly as $R_{vir}\propto(1+z)^{-1}$.  Hence, clusters of fixed mass
have radii that are roughly constant in \emph{comoving} coordinates.}   
Within this
cylinder, we find all galaxies that are connected to the seed galaxy
by the Delaunay mesh (the \emph{first-order Delaunay neighbors}).  If
there are no such galaxies, the seed galaxy is said to be isolated,
and the algorithm moves on to the next seed galaxy in the list.  By
initially searching in a small cylinder, we are able to
limit the probability of chance associations being misidentified as
groups.

If, however, there are one or more first-order Delaunay neighbors, we
move on to Phase II.  We
define a second, larger cylinder, concentric with the first one, with
radius $\mathcal{R}_{II}$ and length $2\mathcal{L}_{II}$.  Within this
cylinder, we identify all galaxies that are connected to the seed
galaxy or to its first-order Delaunay neighbors by the Delaunay mesh.
These are the \emph{second-order Delaunay neighbors}.  The seed
galaxy and its first- and second-order Delaunay neighbors constitute a
set of $N_{II}$ galaxies; we take $N_{II}$ to be an estimate of
the central richness of the group.  Scaling relations are known to
exist between group mass and radius and velocity dispersion
\citep{BN}, and between velocity dipsersion and central richness
\citep{Bahcall}. Thus we may estimate the final size of the group from
$N_{II}$.  In particular, we expect that $N_{II}\propto M \propto
\sigma^3 \propto R^3$.  

Therefore in Phase III we define a third
cylinder, centered on the center of mass of the $N_{II}$ galaxies from 
Phase II, with radius and half-length given by 
\begin{eqnarray} 
\label{eqn:phase3}
\mathcal{R}_{III} &=& r (N_{II}^{corr})^{\frac{1}{3}} \\
\mathcal{L}_{III} &=& \ell (N_{II}^{corr})^{\frac{1}{3}}\nonumber, 
\end{eqnarray}
where $r$ and $\ell$ are free parameters that must be optimized.
Here, the corrected central richness $N_{II}^{corr}$ is scaled to 
account for the redshift-dependent number density $\nu(z)$ of galaxies
in a magnitude-limited survey:
\begin{equation}
N_{II}^{corr} = \left(
\frac{\langle\nu(z)\rangle}{\langle \nu(0.7)\rangle}\right)^{-1} N_{II}.
\label{eqn:nIIcorr}
\end{equation}
We compute $\langle\nu(z) \rangle$ by smoothing the redshift
distribution of the 
entire galaxy sample and dividing it 
by the differential comoving volume element $dV/dz$ to yield the
comoving number density.  All
galaxies within the Phase III cylinder (and any of the $N_{II}$
galaxies from 
Phase II that happen to fall outside of it) are taken to be members of
the group.  After a group has been identified, this three-phase process
repeated on all remaining galaxies that have not yet been assigned to
groups until all galaxies have either been placed into groups or
explicitly identified as isolated galaxies.

The astute reader may object here that we have used the central
richness $N_{II}$ to scale our search window, even though we found
earlier that richness is a relatively unstable group property within
the DEEP2 sample.  This is true.  However, the groups in our mock
catalogs do show some correlation
between actual and observed richness, even though the scatter is
very large.  Furthermore, we have mitigated one major source
of error, Malmquist bias, with the correction in
equation~\ref{eqn:nIIcorr}.  Since the dependence of our scaling on
$N_{II}$ is relatively weak, we anticipate that errors in estimating
this quantity will not introduce insurmountably large errors into our
group sample.  This expectation is borne out by 
tests on mock catalogs, as will be seen in Section~\ref{sec:optimize}.

Finally, it is important to note some minor differences between our
group-finder and the one described in \citet{Marinoni}.  In that
paper, the scaling factors $r$ and $v$ in
Equation~\ref{eqn:phase3} were derived iteratively by running the
group-finder first with a best-guess parameter set and then
automatically adjusting
parameters according to the largest groups found.  In tests on mock
catalogs, we found this method to be unstable, so we instead choose to
optimize our parameters empirically with mock catalogs and then leave
them fixed.  Also, when we search for groups in cylinders, it is
important to note that the ``length'' of our cylinders is supposed to
correspond to an expected maximum velocity of the galaxies in the
group.  Since the mapping between redshift interval and peculiar
velocity changes with redshift, we must rescale the length of our
search cylinders as $\mathcal{L}(z) = [s(z)/s(z_0)]\mathcal{L}_0$,
where the scaling factor $s(z)$ is given by 
\begin{equation}
s(z) = \frac{1+z}{\sqrt{\Omega_M (1+z)^3 + \Omega_\Lambda}}
\label{eqn:zscale}
\end{equation}
for the standard $\Lambda$CDM cosmology.
This scaling amounts to a $\sim 10\%$ effect over the redshift range
of the DEEP2 survey.  We apply it to the cylinder in each phase,
taking a reference redshift of $z_0 = 0.7$.

\subsection{Optimizing with mock catalogs}
\label{sec:optimize}

To gauge the success of our group-finding algorithm, we will make use
of the mock DEEP2 sample described in
Section~\ref{sec:mocks}.   Each galaxy in
these catalogs is tagged with the name of its parent halo, making the
identification of real groups a simple matching exercise.  Thus, we
have a catalog of real groups, identified from N-body models in real
space, against which we can compare the
results of applying the VDM algorithm to the mock galaxy catalog
projected in redshift space.  To
compute completeness and purity, we 
simply apply equation~(\ref{eqn:lgf}) to the real and reconstructed
group catalogs.

As a rough measure of the accuracy of our reconstructed
distribution, $n_{\mathrm{found}}(\sigma,z)$, we apply a
two-dimensional Kol\-mo\-go\-rov-Smirnov (K-S) test to 
this distribution and the real distribution, 
$n_{\mathrm{real}}(\sigma,z)$,  to determine whether they are 
statistically distinguishable. \citet{Marinoni} found that the VDM
group-finder should accurately reproduce this distribution above
$\sigma_c\approx 400\kms$, so we apply the K-S test only above this
velocity dispersion.  The test is insensitive to
the total number of groups in each sample, so we must independently
ensure that the two distributions have the same normalization.  To do
this, we simply count the total number of groups with $\sigma \geq 400
\kms$.  We want to ensure that the
real and reconstructed normalizations match to better than the expected
cosmic variance for our sample (about $12\%$ for the abundance of
groups with $\sigma \ge 400 \kms$), so that our final
errors are dominated by cosmic variance.  Guided by simple physical
considerations (\emph{e.g.}, the expected velocity dispersion range of
groups and clusters), we explore the space of VDM parameters
$\mathcal{R}_\mathrm{min}, \mathcal{L}_\mathrm{min},
\mathcal{R}_\mathrm{II}, \mathcal{L}_\mathrm{II}, r$ and $\ell$, using
trial and error to narrow our parameter range down to a range that
produces an 
$n_{\mathrm{found}}(\sigma,z)$ that is statistically indistinguishable
from $n_{\mathrm{real}}(\sigma,z)$ (less than 1\% confidence that the
two distributions are different) and properly
normalized.  At the same time, we monitor the completeness and
purity of the reconstructed group catalog requiring,
minimally, that $C_2$ and $P_2$ remain above $50\%$ and attempting to
increase them as much as possible.

The procedure described above is simple to implement and perform;
however, it is ultimately an insufficient test of our success.  It
asks only whether or not $n_{\mathrm{found}}(\sigma,z)$ and
$n_{\mathrm{real}}(\sigma,z)$ appear, in a statistical sense, to have
been drawn from the same distribution. But we want to know whether
or not $n_{\mathrm{found}}(\sigma,z)$ is an accurate
\emph{reconstruction} of $n_{\mathrm{real}}(\sigma, z)$; for the two
distributions to pass a K-S 
test is a necessary but not a sufficient condition. In order to fully
optimize our parameters, we must aim to reduce 
any systematic error in $n_{\mathrm{found}}-n_{\mathrm{real}}$ to a
level below the cosmic variance. 

Thus, we will want to assess our error in reconstructing the
acutal velocity function $n_{\mathrm{real}}(\sigma, z)$ in a given
field, irrespective of cosmic variance, and then compare our
reconstruction error to the expected cosmic variance in that field.
As long as the systematic error is smaller than the cosmic variance,
it will not be a significant source of error in our measurement of the
velocity function.   To estimate our systematic error, we apply the VDM
group finder to twelve independent DEEP2 fields and compute the mean
fractional residuals  $\langle\delta_n\rangle\equiv \langle
(n_{\mathrm{found}} -  n_{\mathrm{real}})/n_{\mathrm{found}}\rangle$,
which constitute a measurment of the fractional systematic
reconstruction error. The uncertainty in determining
$\langle\delta_n\rangle$ is then given by the standard error in this
quantity, $\sigma_{\langle\delta\rangle}$.   We have used
\emph{fractional} 
errors here, rather than absolute errors, to distinguish errors in
reconstruction from the intrinsic scatter (cosmic variance) in
$n_{\mathrm{real}}$ and $n_{\mathrm{found}}$.  We can then measure the
fractional cosmic 
variance  (plus Poisson noise) $\sigma_{cos}\equiv (\langle
n_{\mathrm{real}}^2\rangle/\langle n_{\mathrm{real}}\rangle^2 -
1)^{1/2}$ from the mock catalogs and compare it to the systematic
error $\langle\delta_n\rangle$.

\begin{deluxetable}{ccc}
\tabletypesize{\small}
\tablewidth{0pt}
\tablecaption{Parameters used for group finding with the VDM algorithm
in this study.\label{tab:optimal}}
\tablehead{
\colhead{Parameter}\tablenotemark{a} & \colhead{Optimal} & \colhead{High-purity}
}
\startdata
$\mathcal{R}_{min}$ & 0.3   &  0.1  \\
$\mathcal{L}_{min}$ & 7.8   &  5.0  \\
$\mathcal{R}_{II}$  & 0.5   &  0.3  \\
$\mathcal{L}_{II}$  & 6.0   &  5.0  \\
$r$                 & 0.35  &  0.25  \\
$\ell$                 & 14    &  14 
\enddata
\tablenotetext{a}{All values are given in comoving $h^{-1}$ Mpc.}

\end{deluxetable}

\begin{figure}
\centering
\epsfig{width=3.3in,file=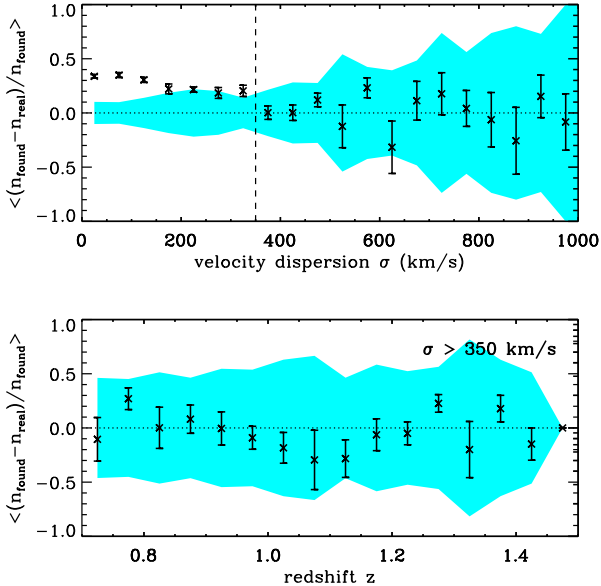}
\caption{Fractional errors in measuring $n(\sigma)$ and $n(z)$.  Upper
  panel: the data points show the fractional systematic error
  $\langle\delta_n\rangle$
  as a function of velocity dispersion,
  estimated by running the VDM algorithm on twelve independent mock
  DEEP2 pointings.  Error bars show the standard deviation of
  the mean $\sigma_{\langle\delta\rangle}$, while the shaded region
(light blue in the electronic edition)
 shows the  fractional cosmic variance 
  (plus Poisson noise) $\sigma_{cos}$ for a single ($120\times 30$
arcmin) DEEP2 field, in bins of $50\kms$.
  For $\sigma \ga 350\kms$, the systematic errors are dominated by
cosmic variance.   
  Bottom panel: Fractional errors in $n_{\mathrm{found}}$ and
  fractional cosmic variance, as a function of redshift in bins of
0.05 in $z$, after 
  groups with $\sigma < 350\kms$ have been discarded.  
  Any systematic offsets are smaller than the cosmic variance.}
\label{fig:group_dist_errs}
\end{figure}

For simplicity of presentation, we first consider the integrated
one-dimensional distributions $n(\sigma)$ and $n(z)$.
Figure~\ref{fig:group_dist_errs} shows the fractional systematic
errors $\langle\delta_n\rangle$ in these distributions, and 
error bars show the uncertainty
$\sigma_{\langle\delta\rangle}$ in determining this quantity.
These two quantities are measured by applying the VDM group-finder to
the DEEP2 mock catalogs using the ``optimal'' VDM parameter
set shown in Table~\ref{tab:optimal}.   Also shown is the fractional
cosmic variance (plus Poisson noise) $\sigma_{cos}$ expected in a
single ($120\times 30$ arcmin) DEEP2 field.
As shown in the upper panel of the figure, systematic reconstruction
errors in  
$n_{\mathrm{found}}(\sigma)$ are dominated by cosmic variance for
$\sigma > 350\;\kms$, while we significantly overestimate the
abundance of lower-dispersion groups.  If we discard the
reconstructed groups with  $\sigma < 350\;\kms$, systematic errors in
the $n_{\mathrm{found}}(z)$ distribution are also smaller than the
cosmic variance, as shown in the bottom panel of the figure.  We note
that we have been able to do somewhat better than expected,
reconstructing the velocity function accurately down to $\sigma =
350\kms$, slightly lower than the cutoff of $400\kms$ expected from
\citet{Marinoni}.  

\begin{figure*}
\centering
\epsfig{width=5.5in,angle=90,file=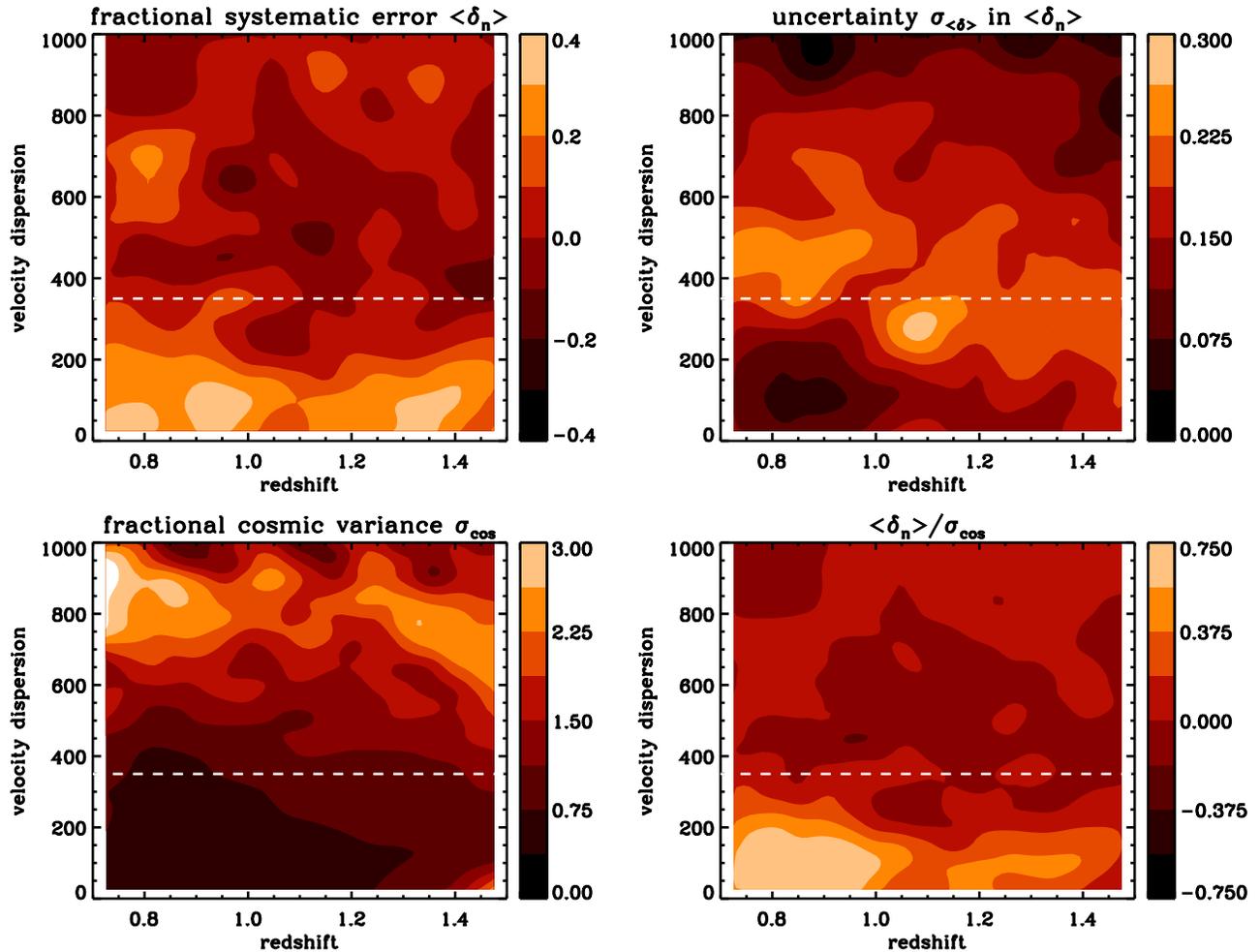}
\caption{Contour plots of the error statistics for the two-dimensional
distribution $n(\sigma,z)$, where the binning in each dimension is the
  same as in 
  figure~\ref{fig:group_dist_errs} and the error distributions have
  been smoothed 
  by one bin in each direction for cleaner presentation.  Upper left:
  The mean fractional systematic  
  error $\langle\delta_n\rangle$ in each bin, computed by applying the
  VDM group finder to twelve 
  independent mock DEEP2 pointings.  Upper right: The standard
  deviation of the mean $\sigma_{\langle\delta\rangle}$ amongst the 12
  mock catalogs for the quantity 
  shown in the upper left panel.  Lower left: The fractional cosmic
  variance (plus Poisson noise) $\sigma_{cos}$
  in each bin, estimated from the same twelve mock pointings.  Lower
  right: The ratio of mean fractional systematic error error to
fractional cosmic 
  variance.  In each panel, contour values are indicated by color
bars. Significant systematic errors in the measured
  distribution are confined to small velocity dispersions, while for 
  $\sigma \ga 350\kms$ the errors are dominated by cosmic variance,
  with no large-scale bin-to-bin correlations.  In each panel, the
  dashed line indicates $\sigma = 350\kms$ for reference.}
\label{fig:group_dist_errs_2d}
\end{figure*}

The fact that these one-dimensional distributions are accurately
measured to within the cosmic variance is heartening, but to fully
optimize our group-finder we must ensure that the full
two-dimensional distribution $n(\sigma, z)$ is accurately measured. 
Figure~\ref{fig:group_dist_errs_2d}
shows smoothed contour plots of the mean fractional systematic errors
$\langle \delta_n(\sigma, z)\rangle$ in this distribution, 
the uncertainty $\sigma_{\langle\delta\rangle})$ in this quantity, the
fractional cosmic variance $\sigma_{cos}$, 
and the ratio of the systematic error to the cosmic variance.  The
panel at the 
lower right shows that significant, correlated overestimates of the
distribution are confined to low velocity dispersion.  For $\sigma \ga
350\kms$, on the other hand, the errors are smaller than the cosmic
variance and exhibit no systematic, large-scale bin-to-bin
correlations.  

To be somewhat more quantitative, we note that for $\sigma \ge 350
\kms$, the average value of the ratio
$(\langle\delta_n\rangle/\sigma_{cos})^2$ 
shown in the lower right panel of Figure~\ref{fig:group_dist_errs_2d}
(before smoothing) is 0.1, and the maximum value is 0.9.  Thus we may
proceed with confidence that $n(\sigma,z)$ for DEEP2 is reconstructed with
sufficient accuracy by the VDM group-finder for velocity dispersions
above $350 \kms$.  
Therefore, having achieved our optimization goals, we may 
apply the VDM algorithm to the DEEP2 redshift catalog with confidence
that our reconstructed catalog will produce an accurate and unbiased
measurement of $n(\sigma,z)$ for $\sigma>350 \kms$ (as long as our
mock catalogs are a reasonable representation of the real universe).
We note that this minimum velocity dispersion is not a limitation of
the VDM group-finder---a more densely sampled survey would permit
$n(\sigma,z)$ to be reconstructed down to even lower dispersions
\citep{Marinoni}.  More generally, it is important to recognize that
the conclusions reached here apply only to the DEEP2 survey: a survey
probing significantly greater volume, for example, would have smaller
cosmic variance, perhaps necessitating a more accurate reconstruction
of $n(\sigma,z)$ than has been presented here.

\begin{figure}
\centering
\epsfig{width=3.3in, file=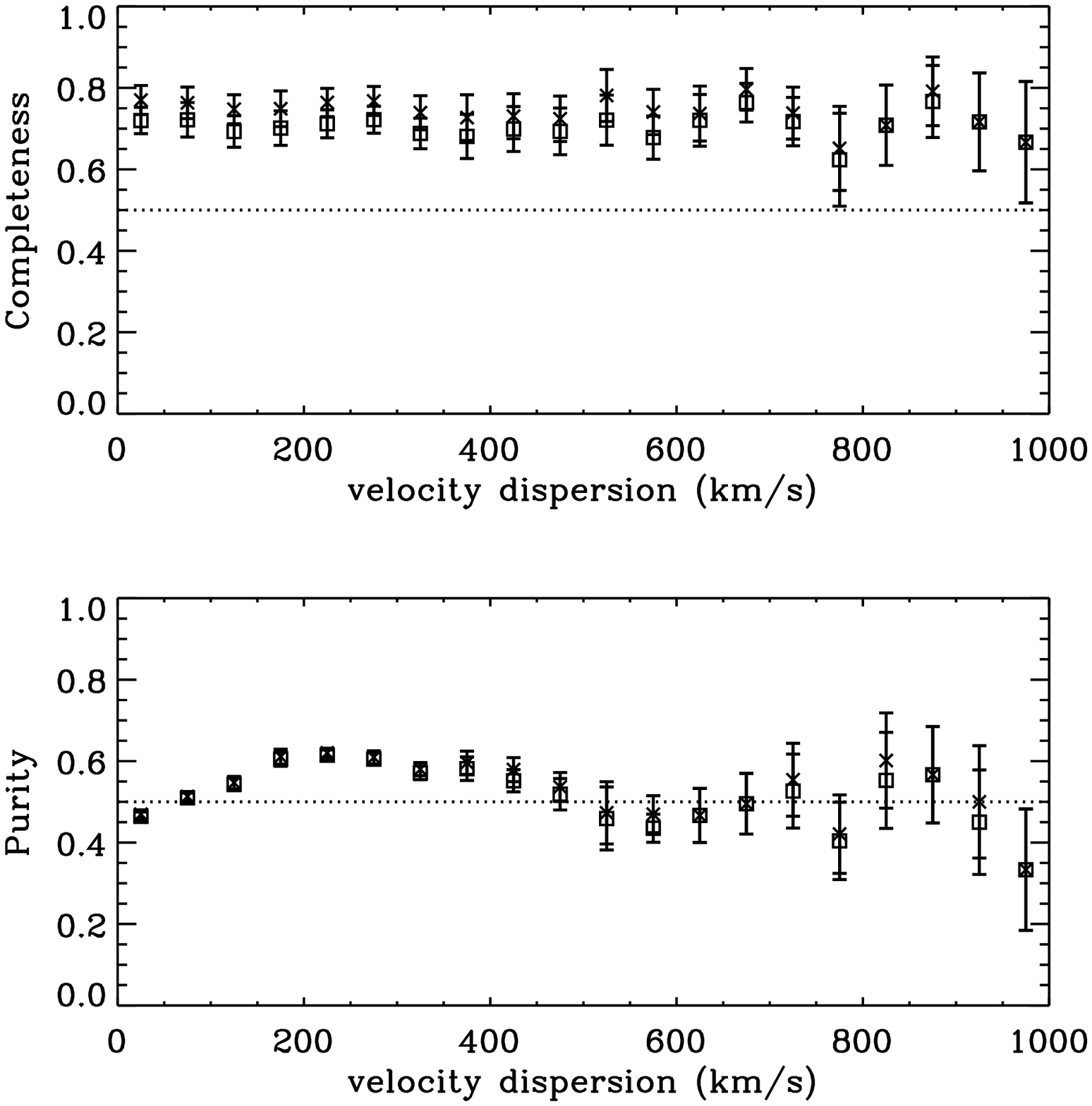}
\caption{Completeness and purity statistics as a function of velocity
dispersion for the optimized VDM
group-finder.  Crosses show the mean one-way statistics $C_1$ and $P_1$
(see Section~\ref{sec:optimal} for definitions), in bins of
$50\;\kms$, as obtained from running
the group-finder on the twelve independent mock catalogs.  Squares
show the mean two-way statistics $C_2$ and $P_2$.  Error bars indicate
the standard deviation of the mean for each statistic in each bin.} 
\label{fig:group_success}
\end{figure}

\begin{figure*}
\centering
\epsfig{width=6in,file=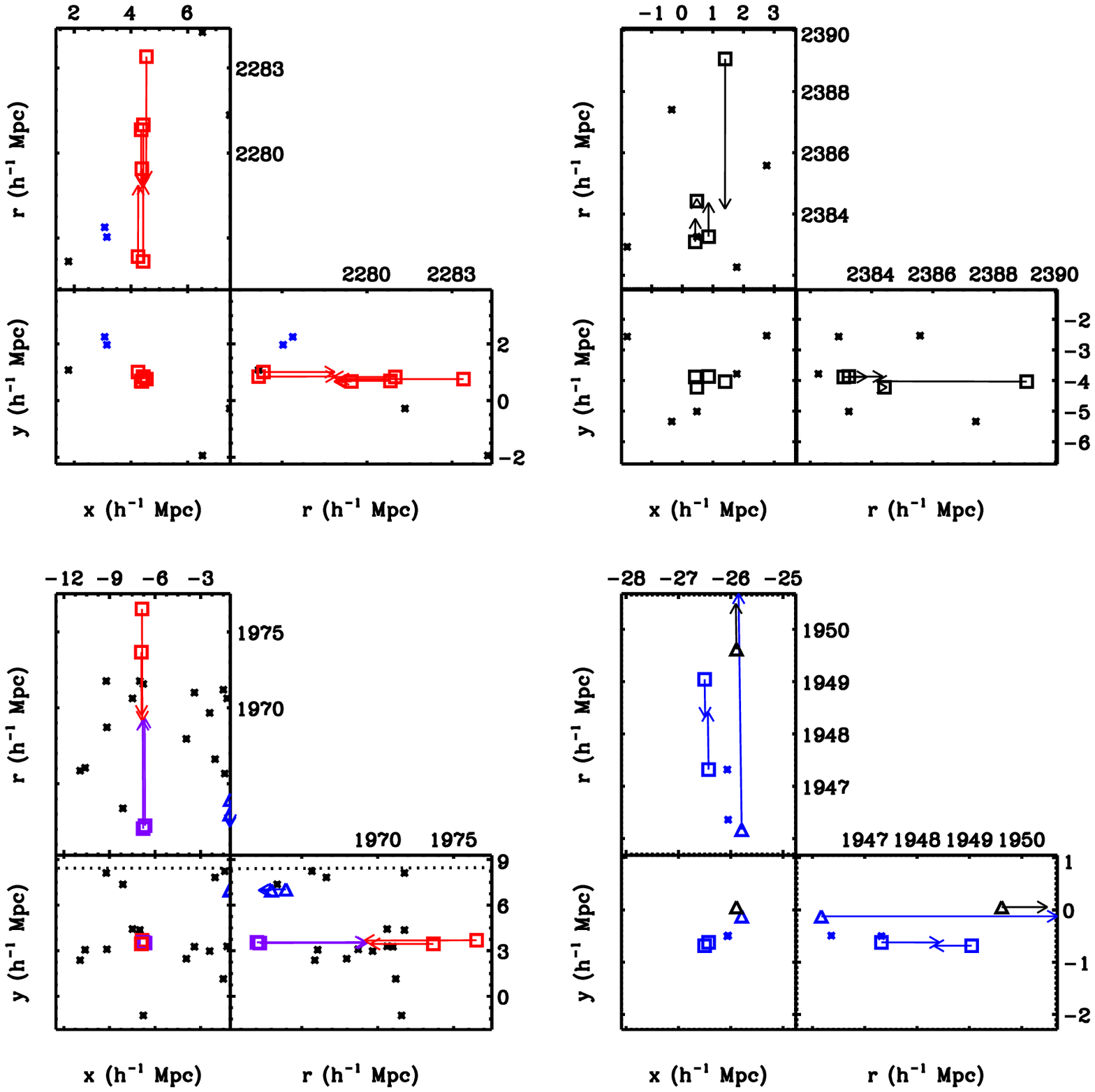}
\caption{Examples of group-finding success and failure in the DEEP2
mock catalogs.  In each panel, squares indicate galaxies in the real
group being plotted, triangles indicate galaxies in nearby real
groups, and crosses indicate nearby field galaxies.  Galaxies are
plotted both as seen on the sky and in twod
line-of-sight projections in redshift space. Arrows point to
the galaxies' real space positions, to show the effects of
redshift-space distortions  
(to reduce visual clutter, no arrows are plotted for field
galaxies). Dotted lines indicate field edges.   
Each reconstructed group is indicated by a different shade (color in
the electronic edition), with
reconstructed field galaxies shown in black.  The upper left panel
shows a perfect reconstruction with a nearby false detection; the
upper right panel shows a completely undetected group (all black
squares); the lower left 
panel shows a fragmented real group; and the lower right panel shows
an example of overmerging.}
\label{fig:closeup_find}
\end{figure*}

After running the VDM on the 12 mock samples using the optimal
parameter set in Table~\ref{tab:optimal}, we  
obtain mean completeness parameters of $C_1 = 0.782\pm 0.006$ and
$C_2=0.719\pm 0.005$ and mean purity parameters of
$P_1=0.545\pm 0.005$ and $P_2=0.538\pm 0.005$, with the quoted
uncertainties indicating the standard deviations of the means.  As
shown in Figure~\ref{fig:group_success}, these statistics are nearly 
independent of the velocity dispersion of the groups being
considered. The fact that $C_1-C_2$ and 
$P_1-P_2$ are small suggests that our catalogs are largely free of
fragmentation or over-merging.  We also find that most galaxies that
belong to real groups are identified as group members: the mocks yield
a mean galaxy-success rate of $S_{Gal}=0.786\pm 0.006$.  Conversely,
the mean interloper fraction is $f_I= 0.458\pm 0.004$, indicating that
the galaxies in our reconstructed group catalogs are dominantly real
group members.

Since the purity is relatively low, it will be difficult to  
know whether to believe in the reality of any \emph{individual} group
in our optimal group sample, although the properties of the catalog as
a \emph{whole} are accurately measured.  To give some sense of the
errors encountered in 
reconstructing individual groups, we show several examples of
group-finding success and failure in Figure~\ref{fig:closeup_find}.
In order to produce a catalog that may be believed with more
confidence on a group-by-group basis, we can optimize the VDM
parameters to maximize the purity (contingent on the requirement that
we still find an appreciable number of groups).  The high-purity
parameter set shown in Table~\ref{tab:optimal} gives mean purity
measures in the mock catalogs of $P_1=0.825\pm 0.007$ and
$P_2=0.815\pm 0.006$. The completeness 
measures are necessarily much lower for this parameter set, however:
$C_1=0.284\pm 0.008$ and $C_2=0.277\pm 0.008$.

%%%%%%%%%%%%%%%%%%%%%  SECTION 5 %%%%%%%%%%%%%%%%%%%%%%%%%%%

\section{The first DEEP2 group catalog}
\label{sec:catalog}

\begin{figure*}[t]
\centering
\epsfig{file=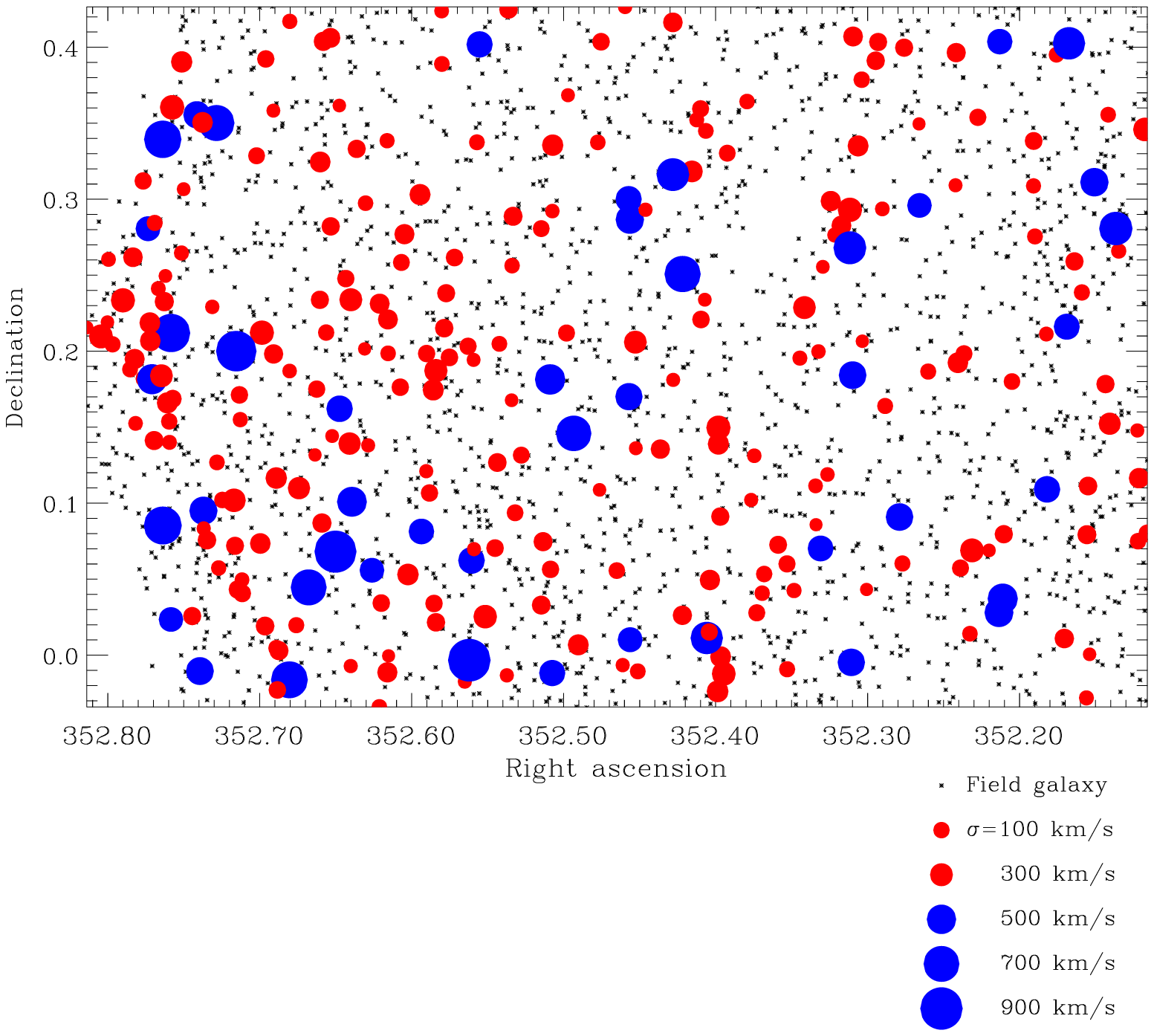, width=5in}
\caption{Groups as seen on the sky in DEEP2 pointing 32 (see
Table~\ref{tab:fields}).  In 
this figure, group-member galaxies have been removed, 
and the median positions of groups are indicated by colored circles
with diameter proportional to measured velocity dispersion, as shown
in the legend.  Groups with $\sigma<350\kms$ are indicated in light
gray (red in the electronic edition),
while groups with $\sigma\ge350\kms$ are shown in dark gray  
(blue).  The positions of field galaxies are indicated by black dots.}
\label{fig:clust_sky}
\end{figure*}

\begin{figure*}
\centering
\epsfig{file=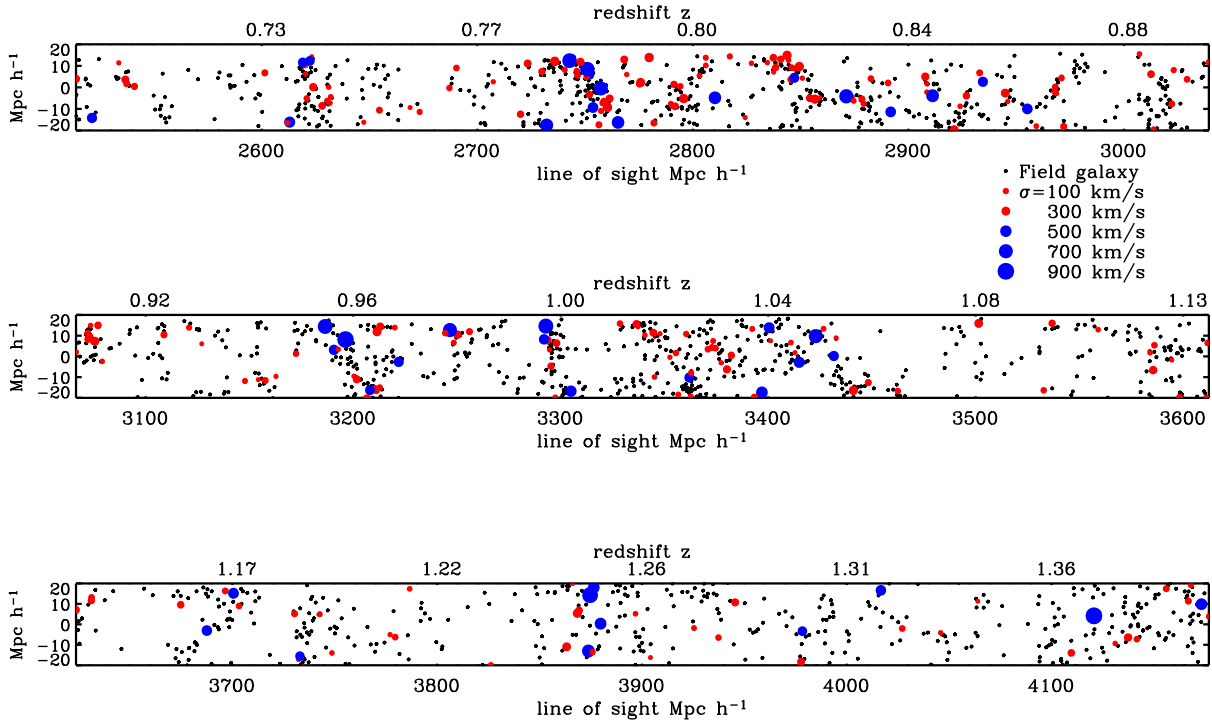,width=4.2in, angle=90}
\caption{Groups of galaxies in DEEP2 pointing 32 as seen along
the line of sight from the observer, projecting through the shortest
dimension of the field.  Symbols are as in Figure~\ref{fig:clust_sky}.  
Line-of-sight distances have been computed using a flat
flat $\Lambda$CDM cosmology with $\Omega_M=0.3$ and equation of state
$w=-1$.  Note that these groups are strongly biased: they
trace out the densest large-scale structures in the survey.}
\label{fig:clust_z}
\end{figure*}

\begin{figure*}[t]
\centering
\epsfig{width=6in, file=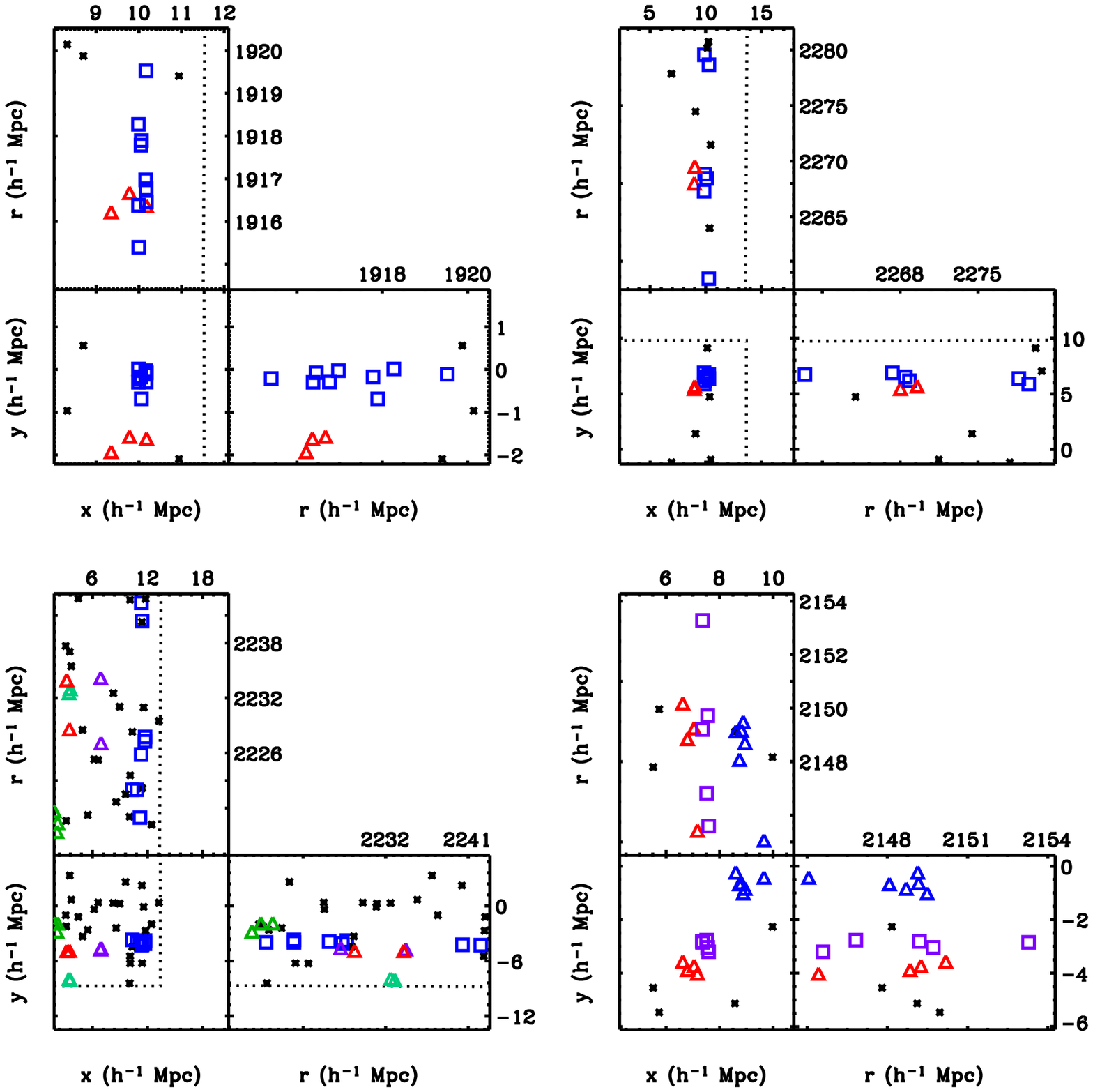}
\caption{Close-up views of four DEEP2 groups.  Shaded squares (colored
in the electronic edition)
indicate galaxies in the group being considered, shaded (colored) triangles
indicate galaxies in nearby groups, and black crosses indicate nearby
field galaxies.  Each group is shown in three projections: one as seen
on the sky and two along the line of sight.}
\label{fig:closeup}
\end{figure*}

We have applied the VDM group finding algorithm to galaxies in the
three 
most completely observed DEEP2 pointings using the optimal parameters
from Table~\ref{tab:optimal}.  Figures~\ref{fig:clust_sky}
and~\ref{fig:clust_z} show groups found in
pointing 32 (see Table~\ref{tab:fields}), both as seen on the
sky, and as seen along the line of sight, projecting 
through the shortest dimension of the field.   
Especially notable in these diagrams is the clear
visual confirmation that groups are strongly biased
tracers of the underlying dark matter distribution.  The groups we
find clearly populate dense regions and filaments preferentially in
these figures.  Close-up
views of a few of the larger groups from this pointing can be seen in
Figure~\ref{fig:closeup}.   

Our optimal VDM group-finder identifies a total of 899 groups with
$N\ge 2$ in the three fields considered here, with $32\%$ of all
galaxies in the sample being placed into groups. We note that this
percentage is much lower than that found in the 2dFGRS by \citet{Eke}
($55\%$); however, our observational selection criteria and
group-finding methods are sufficiently different from theirs that
detailed comparisons will be quite difficult.  By comparing the volume
of the initial search cylinder used in Phase I of the VDM group-finder
to the number density of DEEP2 galaxies in the range $0.7\le z <0.8$,
we estimate that our groups have a  minimum central overdensity
(in redshift space) of $\delta\nu/\nu \ga 100$.   
  
In Table~\ref{tab:groups} we present the locations and properties of
the  subset of groups with
$\sigma \ge 350\kms$ (153 groups). We also have found groups in the
same data 
using the high-purity parameter set in Table~\ref{tab:optimal}.  We
can match our two group catalogs by identifying those groups in the
optimal catalog that are the Largest Associated Groups of the groups
in the high-purity catalog.  Such groups are noted as ``strong''
detections in Table~\ref{tab:groups}; they are highly likely ($>80\%$
chance) to be associated with real virialized structures. Such strong
detections constitute $17\%$ of the total group sample and $13\%$ of
the sample with $\sigma \ge 350 \kms$.

\begin{figure}[h]
\centering
\epsfig{width=3.3in,file=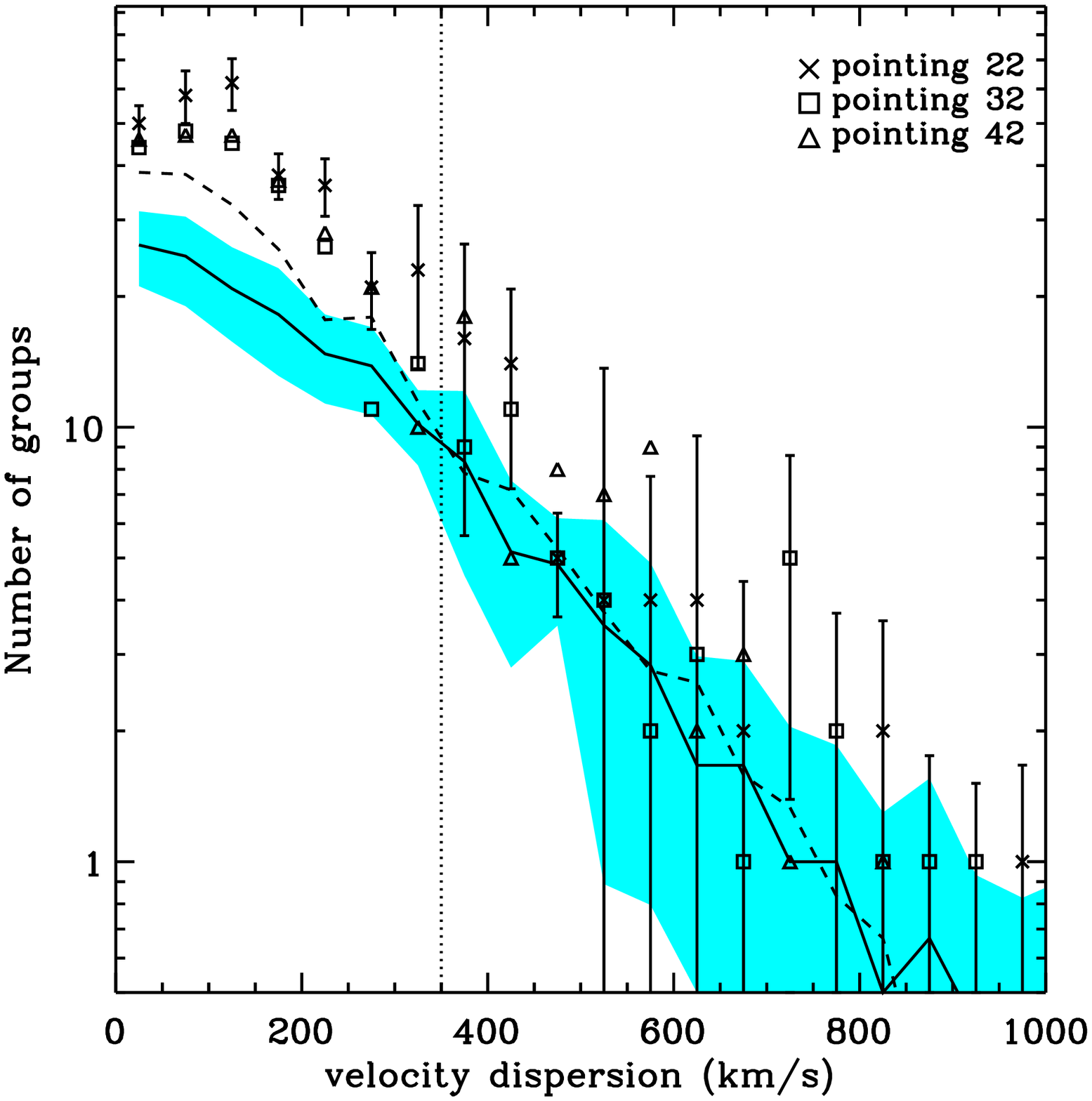}
\caption{Comparison of the velocity function $n(\sigma)$ measured in
a three DEEP2 pointings (see Table~\ref{tab:fields}) to that predicted
by mock catalogs.  The data points are the measured velocity function,
in bins of $50\kms$, of groups found in the three pointings.
Error bars are estimated by applying the VDM
group finder to twelve independent mock DEEP2 pointings, measuring
$n(\sigma)$, and taking the standard deviation of the fractional
residuals $\delta_n$ to estimate a fractional error in each bin. For
clarity, error bars are 
only shown for one data point in each bin. The solid line is the
average ``true'' velocity function 
from the mock DEEP2 catalogs (see Section~\ref{sec:mocks}), $\langle
n_{\mathrm{true}}(\sigma)\rangle$, in bins of $50\kms$, and the
shaded region (light blue in the electronic edition) indicates the
combined cosmic variance and Poisson noise 
in each bin, for a single DEEP2 pointing.  The dashed line shows the
average reconstructed velocity function, $\langle
n_{\mathrm{found}}(\sigma)\rangle$. The measured velocity
function is consistent with the mock catalogs in the regime $\sigma >
350\kms$ (demarcated by the dotted 
line), where an accurate measurement is expected.}
\label{fig:dist_comp}
\end{figure}

Throughout this study, we have focused on reconstructing a group
catalog that provides an accurate measure of the velocity function
$n(\sigma)$.   The ultimate goal of measuring cosmological
parameters must wait for more data, but it is interesting at this
stage to compare the DEEP2 data to the predictions from mock
catalogs.  Figure~\ref{fig:dist_comp} compares the measured velocity
function $n(\sigma)$ (data points) to the true velocity function
$n_{\mathrm{true}}(\sigma)$ (solid line) predicted by the mock DEEP2
sample described in Section~\ref{sec:mocks}.
The measured velocity
functions are qualitatively consistent with the prediction from the mock
catalogs for $\sigma \ga 350\kms$, while the measurements are
significantly higher than the prediction for lower velocity
dispersions.  In Section~\ref{sec:optimize}, we
showed that an accurate reconstruction of the velocity function is
expected in the higher-dispersion regime, while an overestimate of
$n(\sigma)$ is expected at lower velocity dispersions.  We intend to
exclude low-dispersion groups from future analyses, so we do
not consider the high measured values for $n(\sigma)$ a particular
cause for concern.

However, it is important to note that this comparison---of the
measured velocity function to the ``true'' velocity function in the
mock catalogs---is not, strictly speaking, the 
appropriate comparison to make to assess the similarity of the mocks
and the data.  A
real ``apples-to-apples'' comparison would compare the measured
$n(\sigma)$ to the mean \emph{reconstructed} velocity function
$\langle n_{\mathrm{found}}(\sigma)\rangle$ derived by applying the VDM
group-finder to all twelve mocks.  This quantity is
indicated by the dashed line in Figure 13; the data appear to be
consistent with it at velocity dispersions
$\sigma\ga 300\kms$.

Below this threshold a slight discrepancy remains: the data points lie
significantly above the dashed line at low dispersions.
However, one would naively expect the velocity function reconstructed
from the data to be everywhere consistent with the one reconstructed
in the mock catalogs, if the mocks are a good simulation of the data.  
It is difficult to assess the significance of the discrepancy shown
here, since we have not optimized our 
group-finder to measure the abundance of such low-dispersion
groups, but it appears that the the mock catalogs may be 
inconsistent with the DEEP2 data on small velocity scales.   
Nevertheless, it is clear that the mocks are consistent with the data 
in the high-dispersion regime, so 
this comparison  confirms that the mock catalogs are an accurate
simulation of DEEP2 data for our purposes.  The more scientifically
interesting ``data-to-prediction'' comparison shown by the solid line
in Figure~\ref{fig:dist_comp} then stands as evidence that an accurate
reconstruction of the velocity function for $\sigma > 350\kms$ is
possible in DEEP2, providing the first step necessary to placing
constraints on cosmological parameters.

%%%%%%%%%%%%%%%%%%%%%  CONCLUSION %%%%%%%%%%%%%%%%%%%%%%%%%%%%

\section{Discussion and Conclusions}
\label{sec:conclusion}

We have optimized the VDM group-finding algorithm using mock catalogs
designed to replicate the DEEP2 
survey, and we have applied it to spectroscopic data
from three DEEP2 photometric pointings.  In the process of
optimization, 
we have defined a measure of group finding success that focuses on
accurately reproducing the overall properties of the group
catalog---in particular the distribution of groups in redshift and
velocity dispersion, $n(\sigma, z)$---while paying some attention
also to the accurate 
reconstruction of individual groups.  Tests on DEEP2 mock catalogs
show that we are able to accurately reproduce $n(\sigma, z)$ for
$\sigma \ge 350\kms$ and
that errors in measuring this quantity in DEEP2 should be smaller than
its expected intrinsic cosmic variance.  It should thus be possible to
use the test described by \citet{NMCD} to constrain cosmological
parameters, including the dark energy equation of state parameter, $w$. 

We find 899 groups with two or more members  within the DEEP2 data
considered in this study, roughly 
$25\%$ of the expected final sample.  Of these, 153 have 
velocity dispersions $\sigma \ge 350\kms$.  The distribution of
these reconstructed groups with velocity dispersion $n(\sigma)$ is 
in good agreement with the distribution for real groups in
DEEP2 mock catalogs.   This result provides a useful consistency check
for the mock catalogs:  
assuming our reconstructed $n(\sigma)$ is accurate (as our tests
show that it is for $\sigma \ge 350\kms$), we may be confident that
the properties of groups in 
the mock catalogs are an accurate simulation of real DEEP2 data.  This
is especially important in the context of the so-called velocity bias,
which is the ratio of the velocity dispersion of galaxies to that of
the underlying dark matter halo, $b_v =
\sigma_{\mathrm{gal}}/\sigma_{\mathrm{DM}}$.  Various studies of
N-body simulations \citep[\emph{e.g.}, ][and references
therein]{Diemand} have 
suggested that $b_v \ne 1$ at the 15--30\% level, but no such effects
have been included in the DEEP2 mock catalogs.  Our results on
$n(\sigma)$ thus indicate that our data are consistent with $b_v =
1$ within the measurement errors shown in
Figure~\ref{fig:dist_comp}.  This is no surprise, since our 
current error bars are considerably larger than the expected effect;
nevertheless, this result may be viewed as confirmation that no
stronger biases exist.

Successful detection of groups and clusters within the DEEP2 redshift
survey is an essential first step for a wide variety of planned
studies.  By comparing the properties of galaxies in groups to the
properties of isolated galaxies at high redshift, we can learn much
about galaxy formation and evolution.  This will be discussed further
in an upcoming paper~\citep{properties}.  Also, with the catalog of
groups we now have in hand, it is possible to pursue targeted 
follow-up observations in X-rays or using the Sunyaev-Zeldovich
effect to better constrain the gas physics of groups at high redshift;
such programs are now being developed, including an upcoming
X-ray survey of the extended Groth Strip field with the \emph{Chandra}
space telescope.  Finally, using the
groups we find in our current and future spectroscopic data, we expect
to put strong new constraints on the formation and evolution of
galaxies, groups, and clusters, and to investigate the makeup of the
universe, including the nature of the dark energy.

\acknowledgments

We thank Martin White and Peder Norberg for useful discussions about
this project. This work was supported in part by NSF grant 
AST00-71048. BFG acknowledges support from an NSF Graduate Student
Research Fellowship, and JAN is supported by Hubble Fellowship
HST-HF-01165.01-A. The data presented herein were obtained at the
W.M. Keck Observatory, which is operated as a scientific partnership
among the California Institute of Technology, the University of
California and the National Aeronautics and Space Administration. The
Observatory was made possible by the generous financial support of the
W.M. Keck Foundation.  The DEIMOS spectrograph was funded by a grant
from CARA (Keck Observatory), an NSF Facilities and Infrastructure
grant (AST92-2540), the Center for 
Particle Astrophysics and by gifts from Sun Microsystems and the Quantum
Corporation.  The DEEP2 Redshift Survey has been made possible through 
the dedicated efforts of the DEIMOS staff at UC Santa Cruz who built
the instrument and the Keck Observatory staff who have supported it on
the telescope.  Finally, the authors wish to recognize and acknowledge
the very significant cultural role and reverence that the summit of
Mauna Kea has always had within the indigenous Hawaiian community.  We
are most fortunate to have the opportunity to conduct observations
from this mountain.

\pagebreak
\LongTables

\begin{deluxetable}{cccccc}
\tabletypesize{\small}
\tablewidth{0pt}
\tablecaption{Locations and properties of groups with $\sigma \ge
  350\kms$.\label{tab:groups}}  
\tablehead{
\colhead{RA}\tablenotemark{a,b} & 
\colhead{dec}\tablenotemark{a,b}& 
\colhead{$z$}\tablenotemark{b} & 
\colhead{$\sigma$}\tablenotemark{c} &
\colhead{$N$} & \colhead{Strong}\tablenotemark{d} \\
}
\startdata
16 51 15  & +34 53 15  &    0.824  &      530  &        7  &       \\
16 51 54  & +34 52 34  &    0.800  &      390  &        6  &       \\
16 52 27  & +34 49 59  &    0.792  &      540  &        7  &       \\
16 50 59  & +34 50 28  &    0.963  &      410  &        8  &       \\
16 53 09  & +35 00 21  &    0.865  &      620  &        7  &  Y    \\
16 50 36  & +34 49 10  &    0.864  &      610  &        8  &       \\
16 52 53  & +34 55 35  &    0.866  &      420  &        4  &       \\
16 51 49  & +34 51 14  &    0.799  &      350  &        7  &  Y    \\
16 52 10  & +35 03 41  &    0.821  &      360  &        9  &       \\
16 52 49  & +34 59 54  &    0.860  &      560  &        4  &       \\
16 52 26  & +34 46 06  &    0.790  &      390  &        3  &       \\
16 53 10  & +34 59 12  &    0.855  &      400  &        5  &       \\
16 50 02  & +35 05 37  &    0.856  &      480  &        5  &       \\
16 52 13  & +34 57 59  &    0.864  &      350  &        3  &  Y    \\
16 49 53  & +35 06 45  &    0.849  &      810  &        6  &       \\
16 50 04  & +35 08 23  &    0.869  &      360  &        4  &       \\
16 49 59  & +34 50 14  &    0.962  &      380  &        2  &       \\
16 50 56  & +34 58 48  &    0.797  &      400  &        3  &       \\
16 52 08  & +35 02 22  &    0.781  &      360  &        4  &       \\
16 51 31  & +34 55 27  &    0.772  &      510  &        3  &       \\
16 49 53  & +35 08 43  &    0.854  &      440  &        2  &       \\
16 49 51  & +34 50 16  &    0.861  &      420  &        2  &       \\
16 52 50  & +34 52 24  &    1.267  &      660  &        3  &  Y    \\
16 52 25  & +34 57 14  &    0.974  &      850  &        3  &       \\
16 51 26  & +34 57 26  &    1.271  &      380  &        2  &       \\
16 50 21  & +35 01 36  &    1.019  &      480  &        2  &       \\
16 52 48  & +35 04 30  &    1.173  &      420  &        3  &       \\
16 52 28  & +34 51 33  &    0.970  &      510  &        3  &  Y    \\
16 50 26  & +35 08 28  &    1.306  &      450  &        4  &       \\
16 51 02  & +34 59 31  &    0.940  &      400  &        2  &       \\
16 51 35  & +34 49 02  &    0.970  &      390  &        3  &       \\
16 50 38  & +34 45 57  &    1.195  &      450  &        2  &       \\
16 51 15  & +34 51 02  &    0.834  &      450  &        3  &       \\
16 50 29  & +34 43 47  &    0.861  &      420  &        3  &       \\
16 52 04  & +34 52 09  &    1.168  &      370  &        2  &       \\
16 52 12  & +35 05 27  &    0.729  &      430  &        3  &       \\
16 52 22  & +34 49 54  &    1.239  &      580  &        2  &       \\
16 52 20  & +35 07 53  &    0.790  &      410  &        2  &       \\
16 49 58  & +35 00 49  &    0.790  &      360  &        4  &       \\
16 50 34  & +35 05 02  &    0.982  &      580  &        2  &       \\
16 50 18  & +34 48 06  &    1.216  &      380  &        2  &       \\
16 51 35  & +34 56 35  &    0.929  &      570  &        3  &  Y    \\
16 52 16  & +35 09 20  &    0.972  &      600  &        2  &       \\
16 50 02  & +34 47 37  &    1.274  &      420  &        2  &       \\
16 51 58  & +35 08 09  &    0.764  &      660  &        3  &       \\
16 49 52  & +35 05 53  &    1.119  &      970  &        3  &       \\
16 52 58  & +35 02 24  &    0.800  &      400  &        3  &       \\
16 52 25  & +35 10 37  &    0.781  &      420  &        2  &       \\
16 49 48  & +35 09 23  &    0.748  &      600  &        4  &       \\
16 49 54  & +35 09 43  &    1.275  &      460  &        2  &       \\
16 49 48  & +34 51 36  &    0.829  &      370  &        3  &       \\
16 53 08  & +35 00 03  &    0.985  &      360  &        2  &       \\
23 30 30  & +00 03 21  &    0.786  &      360  &        7  &       \\
23 29 14  & +00 11 03  &    0.787  &      450  &        6  &       \\
23 30 40  & +00 02 40  &    1.049  &      700  &        6  &       \\
23 30 35  & +00 09 43  &    0.994  &      410  &        6  &  Y    \\
23 29 19  & +00 04 12  &    1.023  &      390  &        4  &  Y    \\
23 28 36  & +00 18 40  &    0.737  &      470  &        6  &       \\
23 31 02  & +00 12 43  &    0.995  &      770  &        5  &       \\
23 29 49  & +00 17 11  &    1.045  &      470  &        4  &       \\
23 29 58  & +00 08 45  &    0.788  &      690  &        5  &       \\
23 30 56  & +00 05 42  &    1.039  &      460  &        2  &       \\
23 29 06  & +00 05 27  &    0.838  &      440  &        4  &       \\
23 31 05  & +00 16 50  &    0.740  &      360  &        5  &  Y    \\
23 31 03  & +00 05 07  &    0.952  &      760  &        8  &  Y    \\
23 30 57  & +00 21 19  &    1.319  &      430  &        6  &       \\
23 31 03  & +00 20 21  &    0.783  &      750  &        3  &       \\
23 29 41  & +00 15 02  &    0.830  &      710  &        3  &       \\
23 30 22  & +00 04 52  &    0.821  &      380  &        2  &       \\
23 30 51  & +00 12 00  &    1.251  &      840  &        3  &       \\
23 28 40  & +00 12 57  &    1.229  &      400  &        2  &       \\
23 30 02  & +00 10 53  &    1.253  &      530  &        4  &       \\
23 30 14  & +00 03 44  &    0.953  &      420  &        2  &       \\
23 29 14  & -00 00 17  &    0.862  &      440  &        3  &       \\
23 28 51  & +00 24 13  &    0.960  &      380  &        3  &  Y    \\
23 30 54  & +00 21 00  &    0.976  &      700  &        6  &       \\
23 31 05  & +00 10 53  &    1.252  &      520  &        3  &       \\
23 29 49  & +00 18 00  &    0.966  &      400  &        3  &       \\
23 30 36  & +00 04 05  &    0.955  &      890  &        3  &       \\
23 29 03  & +00 17 45  &    1.185  &      350  &        3  &       \\
23 30 33  & +00 06 03  &    1.398  &      510  &        2  &       \\
23 28 50  & +00 02 13  &    1.038  &      530  &        2  &       \\
23 29 14  & +00 16 04  &    1.250  &      610  &        2  &       \\
23 29 42  & +00 18 58  &    0.845  &      600  &        2  &       \\
23 29 49  & +00 10 12  &    1.164  &      430  &        2  &       \\
23 29 49  & +00 00 36  &    1.301  &      360  &        2  &       \\
23 31 00  & +00 17 50  &    1.414  &      500  &        3  &       \\
23 28 43  & +00 06 32  &    0.705  &      410  &        2  &       \\
23 30 13  & +00 24 06  &    0.854  &      400  &        3  &       \\
23 30 43  & -00 00 58  &    0.786  &      730  &        4  &       \\
23 28 51  & +00 01 40  &    1.000  &      490  &        3  &       \\
23 29 37  & +00 00 40  &    0.808  &      590  &        2  &       \\
23 30 14  & -00 00 11  &    1.371  &      910  &        4  &       \\
23 28 32  & +00 16 50  &    0.779  &      620  &        3  &       \\
23 30 01  & -00 00 42  &    1.052  &      400  &        3  &       \\
23 31 02  & +00 01 24  &    0.739  &      350  &        2  &       \\
23 28 40  & +00 24 09  &    0.791  &      590  &        2  &       \\
23 30 57  & -00 00 37  &    1.170  &      450  &        2  &       \\
02 31 21  & +00 35 15  &    0.922  &      580  &        7  &       \\
02 31 13  & +00 34 25  &    0.873  &      620  &       13  &  Y    \\
02 28 58  & +00 40 43  &    0.805  &      550  &        3  &       \\
02 30 33  & +00 27 20  &    0.750  &      520  &        7  &       \\
02 30 25  & +00 37 40  &    0.867  &      480  &        5  &       \\
02 29 26  & +00 47 49  &    0.774  &      690  &        5  &       \\
02 30 15  & +00 41 35  &    0.862  &      490  &        3  &       \\
02 30 44  & +00 44 48  &    0.862  &      360  &        5  &       \\
02 28 42  & +00 26 56  &    0.842  &      820  &        7  &  Y    \\
02 29 57  & +00 32 32  &    0.752  &      500  &        4  &       \\
02 30 15  & +00 46 12  &    0.866  &      480  &        5  &  Y    \\
02 30 21  & +00 26 43  &    0.842  &      560  &        5  &       \\
02 29 37  & +00 30 12  &    0.900  &      390  &        4  &       \\
02 29 27  & +00 28 57  &    0.852  &      530  &        5  &       \\
02 29 06  & +00 34 45  &    0.810  &      380  &        2  &       \\
02 28 58  & +00 27 32  &    0.842  &      410  &        5  &  Y    \\
02 30 29  & +00 34 55  &    0.863  &      350  &        5  &       \\
02 29 13  & +00 29 02  &    0.853  &      440  &        4  &       \\
02 29 56  & +00 36 29  &    0.871  &      720  &        4  &       \\
02 29 53  & +00 34 04  &    0.867  &      390  &        4  &       \\
02 31 02  & +00 36 29  &    0.826  &      640  &        3  &       \\
02 30 06  & +00 34 04  &    0.749  &      420  &        4  &       \\
02 29 31  & +00 30 47  &    0.905  &      370  &        2  &       \\
02 31 12  & +00 42 37  &    0.828  &      580  &        6  &       \\
02 28 58  & +00 44 17  &    1.223  &      600  &        3  &       \\
02 30 05  & +00 45 48  &    0.724  &      360  &        4  &  Y    \\
02 30 59  & +00 37 11  &    0.777  &      380  &        6  &  Y    \\
02 30 41  & +00 28 12  &    0.893  &      570  &        3  &       \\
02 29 18  & +00 25 36  &    0.981  &      380  &        3  &       \\
02 29 36  & +00 32 36  &    0.755  &      670  &        3  &       \\
02 28 51  & +00 33 36  &    1.345  &      360  &        4  &       \\
02 30 06  & +00 36 24  &    0.742  &      580  &        3  &       \\
02 29 31  & +00 36 54  &    1.249  &      350  &        4  &       \\
02 31 20  & +00 32 20  &    0.880  &      680  &        4  &  Y    \\
02 29 24  & +00 42 17  &    1.306  &      490  &        2  &       \\
02 31 03  & +00 43 33  &    1.003  &      580  &        4  &       \\
02 29 04  & +00 30 08  &    1.125  &      460  &        2  &       \\
02 30 37  & +00 41 48  &    0.906  &      350  &        2  &       \\
02 29 46  & +00 31 02  &    1.040  &      360  &        2  &       \\
02 30 37  & +00 45 14  &    0.876  &      360  &        3  &  Y    \\
02 28 47  & +00 36 51  &    1.206  &      510  &        2  &       \\
02 31 00  & +00 34 05  &    1.121  &      400  &        2  &       \\
02 30 39  & +00 43 22  &    1.309  &      350  &        2  &       \\
02 29 16  & +00 31 08  &    0.711  &      570  &        2  &       \\
02 29 49  & +00 30 14  &    1.427  &      360  &        3  &       \\
02 28 50  & +00 38 55  &    0.720  &      450  &        2  &       \\
02 28 41  & +00 38 26  &    0.984  &      400  &        2  &       \\
02 29 20  & +00 24 51  &    0.849  &      520  &        2  &       \\
02 29 53  & +00 43 24  &    0.962  &      530  &        3  &  Y    \\
02 31 12  & +00 29 42  &    0.892  &      400  &        7  &       \\
02 30 16  & +00 48 52  &    1.236  &      530  &        3  &       \\
02 28 38  & +00 30 21  &    0.790  &      470  &        2  &       \\
02 31 23  & +00 34 16  &    0.799  &      540  &        2  &       \\
02 29 59  & +00 21 45  &    1.023  &      370  &        3  &       \\
02 28 44  & +00 47 01  &    0.770  &      440  &        2  &       \\

\enddata

\tablenotetext{a}{Positions on the 
sky are given in J2000 sexagesimal coordinates.}
\tablenotetext{b}{Median value of all galaxies in the group.}
\tablenotetext{c}{Given in $\kms$.}
\tablenotetext{d}{Groups
 detected in both the standard and high-purity 
 group catalogs are indicated as strong detections.} 

\end{deluxetable}

\end{document}